\documentclass{SciPost}

\hypersetup{
    colorlinks,
    linkcolor={red!50!black},
    citecolor={blue!50!black},
    urlcolor={blue!80!black}
}

\usepackage[bitstream-charter]{mathdesign}
\DeclareSymbolFont{usualmathcal}{OMS}{cmsy}{m}{n}
\DeclareSymbolFontAlphabet{\mathcal}{usualmathcal}

\fancypagestyle{SPstyle}{
\fancyhf{}
\lhead{\colorbox{scipostblue}{\bf \color{white} ~SciPost Physics Codebases }}
\rhead{{\bf \color{scipostdeepblue} ~Submission }}

\fancyfoot[C]{\textbf{\thepage}}
}

\usepackage{dcolumn}
\usepackage{bm}
\usepackage{color}
\usepackage{graphicx}
\usepackage{slashed}
\usepackage{pstricks}
\usepackage{float}
\usepackage{array}
\allowdisplaybreaks
\usepackage{dcolumn}
\usepackage{slashed}
\usepackage[inline]{enumitem}
\usepackage{lscape}
\usepackage{isotope}

\makeatletter
\protected\def\xvcenter{%
  \hbox\bgroup$\everyvbox{\everyvbox{}\aftergroup\m@th\aftergroup$\aftergroup\egroup}%
  \vcenter
}
\DeclareRobustCommand{\midscript}[1]{
  \mathchoice{\mid@script\scriptstyle{#1}}
    {\mid@script\scriptstyle{#1}}
    {\mid@script\scriptscriptstyle{#1}}
    {\mid@script\scriptscriptstyle{#1}}
}
\newcommand{\mid@script}[2]{
  \vcenter{\hbox{$\m@th#1#2$}}
}

\DeclareRobustCommand{\textmidscript}[1]{%
  \xvcenter{\hbox{\scriptsize#1}}%
}
\makeatother

\usepackage{color}
\definecolor{gray}{rgb}{0.6,0.6,0.6}
\definecolor{darkgreen}{rgb}{0.0, 0.545098, 0.0}
\definecolor{darkblue}{rgb}{0.0, 0.0, 0.545098}

\mathchardef\mhyphen="2D % hyphen in math mode
\usepackage{siunitx}

\newcommand{\cpp}{C\textmidscript{++}} % prettier C++

\newcommand{\NuHepMCVersion}{1.0.0}

\newcommand{\DiffFlux}{\ensuremath{\phi}}
\newcommand{\IntegratedFlux}{\ensuremath{\Phi}}
\newcommand{\NumInteractions}{\ensuremath{N}}
\newcommand{\TotalXSec}{\ensuremath{\sigma}}
\newcommand{\TargetDensity}{\ensuremath{\rho}}

\newcommand{\AverageFlux}{\ensuremath{\big<\IntegratedFlux\big>}}
\newcommand{\AverageTargets}{\ensuremath{\big<T\big>}}
\newcommand{\AverageXSec}{\ensuremath{\big<\TotalXSec\big>}}
\newcommand{\AverageXSecEstimator}{\ensuremath{\hat{\TotalXSec}}}

\newcommand{\NuSpecies}{\ensuremath{f}}
\newcommand{\NuEnergy}{\ensuremath{E_{\nu}}}
\newcommand{\NuDirection}{\ensuremath{\mathbf{\hat{p}}_\nu}}
\newcommand{\NuTheta}{\ensuremath{\vartheta_\nu}}
\newcommand{\NuPhi}{\ensuremath{\varphi_\nu}}
\newcommand{\NuPosition}{\ensuremath{\mathbf{x}}}
\newcommand{\NuTime}{\ensuremath{t}}

\newcommand{\TargetSpecies}{\ensuremath{j}}

\newcommand{\ProbDist}{\ensuremath{P(\NuSpecies, \TargetSpecies, \NuEnergy,
\NuTheta, \NuPhi, \NuPosition, \NuTime)}}

\newcommand{\NumEvents}{\ensuremath{n}}
\newcommand{\EventIndex}{\ensuremath{e}}

\begin{document}

\pagestyle{SPstyle}

\begin{center}{\Large \textbf{\color{scipostdeepblue}{NuHepMC: A standardized
event record format for neutrino event generators}}}%
\end{center}

\begin{center}\textbf{
% Use (full) first name (+ middle name initials) + surname format.
% Separate subsequent authors by a comma, omit comma and use "and" for the last author.
Steven Gardiner\textsuperscript{1$\star$},
Joshua Isaacson\textsuperscript{1$\dagger$} and
Luke Pickering\textsuperscript{2$\ddagger$}
}\end{center}

\begin{center}
{\bf 1} Fermi National Accelerator Laboratory, P.O. Box 500, Batavia, IL 60510, USA
\\
{\bf 2} STFC, Rutherford Appleton Laboratory, Harwell Oxford, United Kingdom
%%%%%%%%%% END TODO: AFFILIATIONS
% Provide email address of corresponding author(s)
\\[\baselineskip]
$\star$\,\href{mailto:gardiner@fnal.gov}{\small gardiner@fnal.gov}\,,\quad
$\dagger$\,\href{mailto:isaacson@fnal.gov}{\small isaacson@fnal.gov}\,,\quad
$\ddagger$\,\href{mailto:luke.pickering@stfc.ac.uk}{\small luke.pickering@stfc.ac.uk}
\end{center}

\section*{\color{scipostdeepblue}{Abstract}}\textbf{\boldmath%
Simulations of neutrino interactions are playing an increasingly important role
in the pursuit of high-priority measurements for the field of particle physics.
A significant technical barrier for efficient development of these simulations
is the lack of a standard data format for representing individual neutrino
scattering events. We propose and define such a universal format, named NuHepMC,
as a common standard for the output of neutrino event generators. The NuHepMC
format uses data structures and concepts from the HepMC3 event record library
adopted by other subfields of high-energy physics. These are supplemented with
an original set of conventions for generically representing neutrino interaction
physics within the HepMC3 infrastructure.
}

\vspace{\baselineskip}

%%%%%%%%%% BLOCK: Copyright information
% This block will be filled during the proof stage, and finilized just before publication.
% It exists here only as a placeholder, and should not be modified by authors.
\noindent\textcolor{white!90!black}{%
\fbox{\parbox{0.975\linewidth}{%
\textcolor{white!40!black}{\begin{tabular}{lr}%
  \begin{minipage}{0.6\textwidth}%
    {\small Copyright attribution to authors. \newline
    This work is a submission to SciPost Physics Codebases. \newline
    License information to appear upon publication. \newline
    Publication information to appear upon publication.}
  \end{minipage} & \begin{minipage}{0.4\textwidth}
    {\small Received Date \newline Accepted Date \newline Published Date}%
  \end{minipage}
\end{tabular}}
}}
}
%%%%%%%%%% BLOCK: Copyright information

% For convenience during refereeing we turn on line numbers:
%\linenumbers

\vspace{10pt}
\noindent\rule{\textwidth}{1pt}
\tableofcontents
\noindent\rule{\textwidth}{1pt}
\vspace{10pt}

% Write your article contents here, starting from first \section.
% An example structure is given below.

\section{Introduction}
\label{sec:introduction}

Worldwide experimental efforts in high-energy physics (HEP) are placing
increasing emphasis on precision measurements of neutrinos. The pursuit of
these measurements creates strong demands on the quality of neutrino event
generators---the software tools used to simulate neutrino scattering in the
context of experimental analyses~\cite{AlvarezRuso2018,Mosel2019}.

Recent discussions about the future of neutrino research, both at workshops
focused on simulations of neutrino interactions~\cite{Barrow2020} and as part
of the HEP-wide Snowmass 2021 community planning process~\cite{Campbell:2022qmc},
have emphasized the need for greater flexibility in the use of neutrino event
generators and related software. A particularly problematic technical barrier
is the current lack of a common data format for representing the output
\textit{events}: lists of simulated particles involved in a neutrino
interaction together with information describing their properties and
relationships. At present, each neutrino event generator group maintains a
unique output format, substantially complicating the use of multiple generators
in large-scale experimental simulation
workflows~\cite{Aurisano2015,Snider2017}.
The adoption of standard formats in the collider community has streamlined many components of the analysis pipeline. This has enabled straightforward analysis preservation in tools like Rivet~\cite{Buckley:2010ar,Bierlich:2019rhm,Bierlich:2024vqo}, allowed theorists to
reinterpret experimental limits on one exotic physics scenario in light of others~\cite{Butterworth:2016sqg,Buckley:2021neu},
simplified comparisons between generators (see, \textit{e.g.}, Ref.~\cite{Proceedings:2018jsb}),
and supported interoperability between simulation tools~\cite{Alwall:2006yp,Bothmann:2022pwf,Bothmann:2023ozs}.

The only major software product that currently provides an official interface
to the event formats produced by all four of the most widely-used neutrino event
generators (\mbox{GENIE}~\cite{GENIEepj,genie}, GiBUU~\cite{Buss2012},
NEUT~\cite{Hayato:2021heg,Hayato2009}, and NuWro~\cite{Golan2012}) is
NUISANCE~\cite{Stowell2017}, a framework for comparing simulation predictions
to each other and to experimental data. Although this feature of NUISANCE has
proven valuable for the field, its generic internal representation of the input
events (the \texttt{FitEvent} \cpp~class) is too simplified for all
applications, its event format conversion tools require linking to
generator-specific shared libraries, and the need to support multiple evolving
proprietary event formats represents a significant maintenance burden.

To facilitate further development of software products that interface with
neutrino event generators, as well as to support the use of a wider variety of
simulation-based neutrino interaction models in experimental analyses, we
present a new event format, \textit{NuHepMC}, as a universal standard to be
adopted by consensus of the neutrino event generator community. The data
structures, file formats, and basic concepts in NuHepMC are identical to the
versatile and mature HepMC3 library~\cite{Buckley2021} adopted by other
subfields of HEP. Using HepMC3 as a foundation, we define in NuHepMC an
extensible and extendable set of conventions for representing neutrino
interaction physics in a tool-agnostic way. This approach provides a generic
event format that can be used in all future simulation development for neutrino
experiments. Because we avoid placing any limitations on the information that
individual event generators may output, the NuHepMC standard enables lossless,
bidirectional conversion between HepMC3 and the existing proprietary neutrino
event formats.

In Sec.~\ref{sec:spec}, we present the specification of the NuHepMC format.\footnote{For the most up-to-date specification see \url{https://github.com/NuHepMC/Spec}.} The flux-averaged total cross section, a particularly
important quantity for analyzing neutrino scattering simulations, is discussed
in Sec.~\ref{sec:flux_averaged_xsec}. Finally, Sec.~\ref{sec:nuiscomp} uses
NuHepMC as an output format for GENIE, NuWro, NEUT, ACHILLES~\cite{achilles},
and MARLEY~\cite{marleyCPC} as well as an input format in NUISANCE to
demonstrate a first NuHepMC-based analysis of neutrino event generator predictions.

\section{Specification}
\label{sec:spec}

The details of the NuHepMC specification are broken down into three categories
that describe four components from HepMC3. Each element of the specification is
labeled as a \emph{Requirement}, a \emph{Convention}, or a \emph{Suggestion}.
The requirements dictate a minimum level of information to be included when
writing out events. The conventions are optional details that an event
generator group can choose to include or omit while still conforming to the
NuHepMC standard.\footnote{Helper functions and utilities for writing, reading, and interpreting NuHepMC events in \cpp\ or Python can be found at \url{https://github.com/NuHepMC/cpputils}.} Finally, the suggestions are optional recommendations that
are less strongly encouraged than the conventions.

The four HepMC3 components considered in this standard are the generator run
metadata, event metadata, vertex information, and particle information.
Specifications for each of these components can be found in
Secs.~\ref{subsec:run},~\ref{subsec:event},~\ref{subsec:vert},
and~\ref{subsec:part} respectively.

\subsection{Labeling scheme}

The elements of the specification are enumerated below in the form
\\ \mbox{\textless Component\textgreater}.\textless Category\textgreater.\textless
Index\textgreater, where the component of interest is given as G for generator
run metadata, E for event metadata, V for vertex information, and P for
particle information. The category is denoted by R for a requirement, C for a
convention, and S for a suggestion. For example, the second convention for
event metadata would be labeled as \ref{evt:total_xsec}.

If conventions or suggestions prove useful, become widely adopted, and are
considered stable, they may become requirements in future versions of this
specification. \ref{gen:convention} defines a requirement for event generator
authors to signal whether or not specific optional elements of this
specification have been followed.

\subsection{HepMC3 C++ classes}
\label{subsec:hepmc3c++}

Throughout this standard, references are made to various HepMC3 \cpp\ classes,
\textit{e.g.},\\ \texttt{HepMC3::GenRunInfo}. However, these are used as a convenient
handle for data objects. This specification should not be considered specific
to the HepMC3 \cpp\ reference implementation.

In principle, the HepMC3 ASCII format can be written out to file without any required dependencies. However, it is important to note that the HepMC3 library does provide bindings for both \cpp\ and Python. The implementation within each tool using the NuHepMC format is left to the developers. At the time of writing there are official or unofficial conversion tools available for all of the aforementioned neutrino event generators, see Section~\ref{sec:nuiscomp}.

\subsection{Generator Run Metadata}
\label{subsec:run}

The generator run metadata describes the overall setup of the event generator,
\textit{i.e.}, information that is not unique to a specific event. The NuHepMC
specifications for this metadata are as follows:

\subsubsection{Requirements}
\begin{enumerate}[label={\textsc{G.R.\arabic*}},leftmargin=3.5\parindent]

% G.R.1
\item \textsc{Valid GenRunInfo}:\\ \label{gen:run_info}\noindent
All NuHepMC vectors\footnote{The term \textit{vector} is used herein, rather than \textit{file}, as HepMC3 events are frequently piped or streamed between MC tools without ever being wholly persisted to memory or disk. } must contain a \texttt{HepMC3::GenRunInfo} instance.

% G.R.2
\item \textsc{NuHepMC Version}:\\ \label{gen:version}\noindent
A NuHepMC \texttt{HepMC3::GenRunInfo} instance must contain the following three \\ \texttt{HepMC3::IntAttribute}s that specify the version of NuHepMC that is implemented:
\begin{itemize}
    \item \texttt{"NuHepMC.Version.Major"}
    \item \texttt{"NuHepMC.Version.Minor"}
    \item \texttt{"NuHepMC.Version.Patch"}
\end{itemize}

This document describes \textbf{version~\NuHepMCVersion}~of NuHepMC.\footnote{A script is provided for converting existing NuHepMC files from an older version of the standard to the latest version of the standard at \url{https://github.com/NuHepMC/Spec/scripts/NuHepMC_update_standard.py}}

% G.R.3
\item \textsc{Generator Identification}:\\ \label{gen:gen_id}\noindent
A NuHepMC \texttt{HepMC3::GenRunInfo} instance must contain a \\\texttt{HepMC3::GenRunInfo::ToolInfo} for each `tool' involved in the production of the vector thus far. The \texttt{ToolInfo} instance must contain non-empty name and version fields.

% G.R.4
\item \textsc{Signaling Followed Conventions}:\\ \label{gen:convention}\noindent
To signal to a user that an implementation follows a named convention from
this specification, a \texttt{HepMC3::VectorStringAttribute} attribute must be added to
the \texttt{HepMC3::GenRunInfo} instance named \texttt{"NuHepMC.Conventions"}
containing the labels of the conventions adhered to, \textit{e.g.} \texttt{"G.C.2"}. \textit{n.b.} that \ref{gen:fatx} requires the implementation of at least one convention and so this attribute must exist.

% G.R.5
\item \textsc{Flux-Averaged Total Cross Section}:\\ \label{gen:fatx}\noindent
The flux-averaged total cross section, \AverageXSec, is a scaling factor that is
needed to convert a distribution of simulated events into a cross-section prediction. Details on
the mathematical definition of this quantity are given in
Sec.~\ref{sec:av_xsec_def}. If the value of \AverageXSec\ is known at the start of a generator run, convention~\ref{gen:fatcs} should be used to store it in the \texttt{HepMC3::GenRunInfo}.

In general, the value of \AverageXSec\ is not known at the start of a generator run, but it can be calculated as events are produced (see Appendix~\ref{app:av_xsec_mc}).
In this case, convention~\ref{evt:est_xsec} should be used to store the
running estimate and associated statistical uncertainty in each \texttt{HepMC3::GenEvent}.

One of these two methods for communicating \AverageXSec\ to users must be implemented and signaled via \ref{gen:convention}.

% G.R.6
\item \textsc{Cross Section Units and Target Scaling}:\\ \label{gen:cs_units}\noindent
There are a variety of units typically used to report both measured and
predicted cross sections in HEP. For neutrino cross sections specifically,
\SI{e-38}{\centi\meter\squared} per nucleon is common, but not ubiquitous. Both of the
following \\ \texttt{HepMC3::StringAttribute}s must be included on the
\texttt{HepMC3::GenRunInfo} to indicate the cross section units used within a
vector. Possible values of the
attributes are not restricted by this specification, but the meanings of the following values are reserved to standardize existing conventions. It is strongly recommended that new implementations use these wherever possible.

\begin{itemize}
\item \texttt{"NuHepMC.Units.CrossSection.Unit"}:
    \begin{itemize}
	\item \texttt{"pb"}: Picobarns or \SI{e-36}{\centi\meter\squared}. Our recommendation.
	\item \texttt{"1e-38 cm2"}: The choice of
\SI{e-38}{\centi\meter\squared} is the most frequent in the
neutrino literature.
	\item \texttt{"cm2"}: Using bare \si{\centi\meter\squared} in this
option, without any power-of-ten scaling, is not recommended due to numerical
precision concerns. The natural scale of neutrino--nucleon cross sections is
approximately \num{e-38}, which is very close to the minimum representable IEEE
754 single-precision floating point number~\cite{8766229}.
    \end{itemize}
\item \texttt{"NuHepMC.Units.CrossSection.TargetScale"}:
    \begin{itemize}
	\item \texttt{"PerAtom"}: Our recommendation. For this choice, the target number density $\TargetDensity$ mentioned in Sec.~\ref{sec:flux_averaged_xsec} is expressed in atoms per unit volume.
    \item \texttt{"PerNucleon"}: A common alternative in the literature.
    \end{itemize}
\end{itemize}

The chosen units should be assumed to apply to the reported value of \AverageXSec, which is a property of the generator run and may be stored according to either \ref{gen:fatcs} or \ref{evt:est_xsec}. Consistent units should also be used for the total (\ref{evt:total_xsec}) and process-specific (\ref{evt:proc_xsec}) cross sections describing the primary interaction in each individual event.

It is ultimately up to the user to parse these attributes and decide whether
any additional scaling is needed for their purposes, but the use of the above reserved values will facilitate automated processing. The recommended value of picobarns per atom is chosen to remain consistent with the assumptions of other
tools that read and write HepMC3 files beyond the neutrino community, such as Rivet~\cite{Buckley:2010ar,Bierlich:2019rhm,Bierlich:2024vqo}.
Consistency facilitates interoperability and reduces maintenance burdens.

% G.R.7
\item \textsc{Event Weights}:\\ \label{gen:evt_wgt}\noindent
For weights that will be calculated for every event, HepMC3 provides an
interface for storing the weight names only once in the \texttt{HepMC3::GenRunInfo}
instance. At least one event weight named \texttt{"CV"} must be declared in the
\\ \texttt{HepMC3::GenRunInfo} instance and filled for every event.

This weight may be \num{1} or constant for every event in a generator run (in the
case of an \emph{unweighted} event vector). This weight must always be included
by a user when producing cross section predictions from a NuHepMC vector
and should never be assumed to be \num{1} for every event.

% G.R.8
\item \textsc{Process Metadata}:\\ \label{gen:proc_id}\noindent
A NuHepMC \texttt{HepMC3::GenRunInfo} instance must contain a \\ \texttt{HepMC3::VectorIntAttribute} named \texttt{"NuHepMC.ProcessIDs"} listing all physics process IDs as integers. For each valid process ID, the \\ \texttt{HepMC3::GenRunInfo} instance must also contain two other attributes giving a name and description of each:
\begin{itemize}
    \item type: \texttt{HepMC3::StringAttribute},\\ name: \texttt{"NuHepMC.ProcessInfo[$<$ID$>$].Name"}
    \item type: \texttt{HepMC3::StringAttribute},\\ name: \texttt{"NuHepMC.ProcessInfo[$<$ID$>$].Description"}
\end{itemize}
where \texttt{$<$ID$>$} enumerates all process IDs present in \texttt{"NuHepMC.ProcessIDs"}. (See also~\ref{evt:proc_id}).

% G.R.9
\item \textsc{Vertex Status Metadata}:\\ \label{gen:vert_status}\noindent
The NuHepMC \texttt{HepMC3::GenRunInfo} instance must contain a \\ \texttt{HepMC3::VectorIntAttribute} named \\ \texttt{"NuHepMC.VertexStatusIDs"} declaring any generator-specific status codes used. Including the standard HepMC3 codes in this list is optional, but they must not be reused to mean something different than in the HepMC3 specification. For each declared vertex status, the \texttt{HepMC3::GenRunInfo} instance must also contain two other attributes giving a name and description of each:
\begin{itemize}
    \item type: \texttt{HepMC3::StringAttribute},\\ name: \texttt{"NuHepMC.VertexStatusInfo[$<$ID$>$].Name"}
    \item type: \texttt{HepMC3::StringAttribute},\\ name: \texttt{"NuHepMC.VertexStatusInfo[$<$ID$>$].Description"}
\end{itemize}
where \texttt{$<$ID$>$} enumerates all status codes present in \\ \texttt{"NuHepMC.VertexStatusIDs"} (See also~\ref{vert:status}).

% G.R.10
\item \textsc{Particle Status Metadata}:\\ \label{gen:part_status}\noindent
The NuHepMC \texttt{HepMC3::GenRunInfo} instance must contain a \\ \texttt{HepMC3::VectorIntAttribute} named \\ \texttt{"NuHepMC.ParticleStatusIDs"} declaring any generator-specific status codes used. Including the standard HepMC3 codes in this list is optional, but they must not be reused to mean something different than in the HepMC3 specification. For each valid particle status, the \texttt{HepMC3::GenRunInfo} instance must also contain two other attributes giving a name and description of each:
\begin{itemize}
    \item type: \texttt{HepMC3::StringAttribute},\\ name: \texttt{"NuHepMC.ParticleStatusInfo[$<$ID$>$].Name"}
    \item type: \texttt{HepMC3::StringAttribute},\\ name: \texttt{"NuHepMC.ParticleStatusInfo[$<$ID$>$].Description"}
\end{itemize}
where \texttt{$<$ID$>$} enumerates all status codes present in \\ \texttt{"NuHepMC.ParticleStatusIDs"} (see~\ref{part:status} for more details).

% G.R.11
\item \textsc{Non-standard Particle Numbers (PDG Codes)}: \\ \label{gen:pdg_nums}\noindent
Essentially all event generators in HEP use a standard set of integer codes
to identify particle species. This numbering scheme is maintained by
the Particle Data Group (PDG) and is regularly updated in their Review
of Particle Physics~\cite[Sec.~45, p.~733]{PDG}.

It is expected that neutrino event generators may need to use codes for non-standard
particle species (\textit{i.e.}, those without an existing PDG code) for a variety of
applications. This could include simulating exotic physics processes involving
new particles as well as implementing bookkeeping methods involving
generator-specific pseudoparticles.

The NuHepMC \texttt{HepMC3::GenRunInfo} instance must contain a
\\\texttt{HepMC3::VectorIntAttribute} named \\
\texttt{"NuHepMC.AdditionalParticleNumbers"} declaring any particle codes used
that are not defined in the current PDG numbering scheme. Including any of the
standard codes in this list is permitted but not required. The standard
particle codes must not be reused to mean something different than in the PDG
specification.

For each additional particle code, the \texttt{HepMC3::GenRunInfo} instance
must also contain an attribute giving a unique name to the represented particle
species:

\begin{itemize}
    \item type: \texttt{HepMC3::StringAttribute},\\ name: \texttt{"NuHepMC.AdditionalParticleInfo[$<$PDG$>$].Name"}
\end{itemize}
where \texttt{$<$PDG$>$} enumerates all particle numbers present in \\\texttt{"NuHepMC.AdditionalParticleNumbers"}.

See also \ref{gen:nonstdpdg_desc} for a suggested way of storing descriptions
of these special particle species.

\end{enumerate}

\subsubsection{Conventions}
\begin{enumerate}[label={\textsc{G.C.\arabic*}},leftmargin=3.5\parindent]

% G.C.1
\item \textsc{Vector Exposure}:\\ \label{gen:exposure_exp}\noindent
Each vector should contain a description of the exposure of the generator
run. When simulating with some experimental exposure, often represented for
accelerator neutrino experiments in units of ``protons on target'' (POT), the
exposure should be described. Two attributes are reserved for signaling the
exposure to users. One or both can be provided.
\begin{itemize}
    \item type: \texttt{HepMC3::DoubleAttribute}, name: \texttt{"NuHepMC.Exposure.POT"}
    \item type: \texttt{HepMC3::DoubleAttribute}, \\ name: \texttt{"NuHepMC.Exposure.Livetime"}
\end{itemize}

% G.C.2
\item \textsc{Flux-averaged Total Cross Section}:\\ \label{gen:fatcs}\noindent
 If known at the start of a
run, the value of \AverageXSec\ should be stored as a
\\ \texttt{HepMC3::DoubleAttribute} in the \texttt{HepMC3::GenRunInfo} named
\\ \texttt{"NuHepMC.FluxAveragedTotalCrossSection"}. Optionally, the uncertainty in the flux-averaged total cross section may be stored as a \\ \texttt{HepMC3::DoubleAttribute} in the \texttt{HepMC3::GenRunInfo} named \\ \texttt{"NuHepMC.FluxAveragedTotalCrossSectionUncertainty"}.

See~\ref{evt:est_xsec} if the cross section is not known at the start. Also, recall that from~\ref{gen:fatx}, either this convention or~\ref{evt:est_xsec} must be used.

% G.C.3
\item \textsc{Citation Metadata}:\\ \label{gen:citations}\noindent
Modeling components implemented based on published work should always be fully cited. The \texttt{HepMC3::GenRunInfo} should contain at least one \\\texttt{HepMC3::VectorStringAttribute} for each relevant modeling component, named according to the pattern \texttt{"NuHepMC.Citations.$<$Comp$>$.$<$Type$>$"}. Valid substitutions for the \texttt{$<$Comp$>$} and \texttt{$<$Type$>$} fields are not restricted by this standard beyond the requirement that they are pure mixed-case alpha-numeric. The keys \texttt{$<$Comp$>$=Generator} and \texttt{$<$Comp$>$=Process[$<$ID$>$]} are reserved for use in specifying the main citation(s) for the interaction generator and for citing theoretical motivations behind individual processes, respectively. For common reference formats in the HEP field, some reserved values for the \texttt{$<$Type$>$} field are:
\begin{itemize}
     \item \texttt{"InspireHep"} might contain one or more unique InspireHep identifiers (texkeys).
     \item \texttt{"arXiv"} might contain one or more unique arXiv identifiers (eprint numbers).
     \item \texttt{"DOI"} might contain one or more unique Digital Object \mbox{Identifiers}.
\end{itemize}

A tool that can read this metadata and automatically produce a BibTeX file containing entries for all cited publications is briefly introduced in Appendix~\ref{appendix:bib}.

% G.C.4
\item \textsc{Beam Energy Distribution Description}:\\ \label{gen:beam_energy}\noindent
Each vector should contain a description of the incident flux of beam particles used to simulate the events. For many MC studies and experimental simulations in which the detector is not physically close to the source, a simple univariate energy distribution is enough to describe the particle beam. The two types of energy distribution covered by this convention are mono-energetic beams and those with energy spectra that can be described by one-dimensional histograms. The type should be signaled via a \texttt{HepMC3::StringAttribute} named \texttt{"NuHepMC.Beam.Type"} with value \\ \texttt{"MonoEnergetic"} or \texttt{"Histogram"} stored on the \texttt{HepMC3::GenRunInfo}. For both types, relevant units can be signaled via two attributes:
\begin{itemize}
    \item\texttt{"NuHepMC.Beam.EnergyUnit"}. Possible values of the attribute are not restricted, but the meanings of \texttt{"MEV"} and \texttt{"GEV"} are reserved. This attribute should always exist and be not empty.
    \item\texttt{"NuHepMC.Beam.FluxUnit"}. Possible values of the attribute are not restricted, but the meaning of \texttt{"Arbitrary"} is reserved to signal that the normalization of the distribution was not known or used by the simulation. If this attribute is not used then the normalization will be assumed to be arbitrary.
\end{itemize}

For the case of a \texttt{"MonoEnergetic"}-type distribution, all beam particles in the vector must have identical energy. The attribute \\\texttt{"NuHepMC.Beam[$<$PDG$>$].MonoEnergetic.Energy"} can be used to signal the beam energy in the lab frame, but the usage of this attribute is optional as the energy can be determined from the first (or any) event in the vector.

For the case of a \texttt{"Histogram"}-type distribution, the histogram should be encoded into two \texttt{HepMC3::VectorDoubleAttribute} per beam species on the \\ \texttt{HepMC3::GenRunInfo}:
\begin{itemize}
    \item\texttt{"NuHepMC.Beam[$<$PDG$>$].Histogram.BinEdges"}
    \item\texttt{"NuHepMC.Beam[$<$PDG$>$].Histogram.BinContent"}
\end{itemize}

where \texttt{$<$PDG$>$} enumerates the PDG particle numbers of all beam particles present in the event vector. \textit{n.b.} the number of entries in the \texttt{"BinEdges"} vector should always be one more than the number of entries in the \texttt{"BinContent"} vector.

The \texttt{HepMC3::BoolAttribute},
\begin{itemize}
    \item\texttt{"NuHepMC.Beam[$<$PDG$>$].Histogram.ContentIsPerWidth"},
\end{itemize}
should be set to \texttt{true} to indicate that the total neutrino flux in a bin of the histogram should be computed by multiplying the bin content by the bin width. A value of \texttt{false}, assumed by default when this attribute is not present, signals that the bin content should be used alone. While this distinction might be determined by carefully parsing the \texttt{FluxUnit} attribute, existing neutrino generators make different assumptions when sampling from input neutrino beam energy distributions. An explicit attribute is therefore defined here to reduce ambiguity.

For a suggestion on how to encode additional information about more realistic neutrino beam descriptions, see \ref{evt:beam_description}.

General flux handling is a complex problem that is not currently standardized. This convention is suitable for simplified studies by theorists and phenomenologists interested in accelerator-based neutrino experiments. A standardized flux format usable for many other applications, \textit{e.g.}, atmospheric neutrino measurements, as well as its corresponding description within the event record, is outside the scope of this specification. Development of such a standard is left to future community efforts.

% G.C.5
\item \textsc{Non-standard Particle Number Descriptions}:\\ \label{gen:nonstdpdg_desc}\noindent
For each additional particle number \texttt{$<$PDG$>$} declared in the
\\\texttt{"NuHepMC.AdditionalParticleNumbers"} attribute, according to
\ref{gen:pdg_nums}, the \texttt{HepMC3::GenRunInfo} instance may contain an
attribute giving a description of the particle:
\begin{itemize}
    \item type: \texttt{HepMC3::StringAttribute},\\ name:
\texttt{"NuHepMC.AdditionalParticleInfo[$<$PDG$>$].Description"}
\end{itemize}

For non-standard particles that should be further simulated by particle propagation simulations, such as GEANT4~\cite{GEANT4:2002zbu}, additional information encoded here may be enough to enable automatic propagation. In this version of NuHepMC, no attempt is made to standardize a format for such information, but it is suggested that \texttt{HepMC3::GenRunInfo} attributes of the form, \\\texttt{"NuHepMC.AdditionalParticleNumber[$<$PDG$>$].$<$SimName$>$.$<$AttrName$>$"}, may be useful and could be standardized in future versions of this specification. These additional attributes could be expected to include, at minimum,
the particle's mass, width, spin, and electric charge.

\end{enumerate}

\subsubsection{Suggestions}
\begin{enumerate}[label={\textsc{G.S.\arabic*}},leftmargin=3.5\parindent]

% G.S.1
\item \textsc{Run Configuration}:\\ \label{gen:run_conf}\noindent
It is suggested that a NuHepMC \texttt{HepMC3::GenRunInfo} instance should contain all information required to reproduce the events in the vector. This may be stored in attributes with names beginning with \texttt{"NuHepMC.Provenance"}. The information required will necessarily be generator-specific, but it should be noted that storing the full initial state of any random number generators used during the generator run is crucial for exact reproduction.

% G.S.2
\item \textsc{Complete Status Metadata}:\\ \label{gen:complete_status_meta}\noindent
While \ref{gen:vert_status} and \ref{gen:part_status} explicitly do not require implementations to emit metadata for standard status codes defined in the HepMC3 standard, it is suggested that the complete list of status codes used by an implementation are included in the \texttt{"NuHepMC.VertexStatusInfo"} and \texttt{"NuHepMC.ParticleStatusInfo"} attributes.

% G.S.3
\item \textsc{Target Atomic Abundances}:\\ \label{gen:fatm}\noindent
Using the flux-averaged total cross section \AverageXSec, it is straightforward to compute partial cross sections for specific final states, interaction modes, and components of the overall target material used in a run. An example of the last of these is a flux-averaged cross section for interactions with $\textrm{C}$ when a composite $\textrm{CH}$ target was simulated.

In some cases, it can be desirable to use an existing vector to create cross-section predictions for a target material with a different composition than the one initially used, \textit{e.g.}, distributions for $\textrm{CH}_2$ from a vector generated for a $\textrm{CH}$ target. Making such predictions requires a knowledge of the flux-averaged relative abundances of the nuclides present in the original target material. For situations in which these are known at the start of a run, a
\texttt{HepMC3::VectorStringAttribute} named
\texttt{"NuHepMC.TargetMaterialRelativeAtomicAbundance"}
may be stored in the \texttt{HepMC3::GenRunInfo}. Each element
of the vector should be a string of the form \texttt{"100ZZZAAA0[x]"} containing a nuclear PDG code and the corresponding relative abundance enclosed in square brackets []. For nuclides with a relative abundance of \num{1}, the nuclear PDG code may optionally be used alone. For example, a pure $\textrm{CH}_2$ target might be represented in this format as \texttt{["1000060120[0.5]", "1000010010"]}.

For realistic experimental simulations that involve a detector and a non-uniform neutrino beam, the flux-averaged relative abundances are typically difficult to calculate analytically. While running estimates of the abundances can in principle be obtained as the generator runs (using techniques akin to those in~\ref{evt:est_xsec}), alternative approaches will generally be simpler to implement. Cases in which a particular detector material is of interest can be addressed by selecting events based on their position within the lab frame (\textit{c.f.} \ref{evt:lab_pos}). Further exploration of this topic is left to future standardization efforts focused on neutrino fluxes and detector geometries.

\end{enumerate}

\subsection{Event Data}
\label{subsec:event}

The event is used to store information about one primary interaction process and any relevant secondary processes. An event is described by arbitrary metadata and a graph of particles (edges) and vertices (nodes), each with their own arbitrary metadata.
The NuHepMC specifications for events are as follows:

\subsubsection{Requirements}
\begin{enumerate}[label={\textsc{E.R.\arabic*}},leftmargin=3.5\parindent]

% E.R.1
\item \textsc{HepMC3 Compatibility}:\\ \label{evt:hepmc3}\noindent
The HepMC3 standard places some constraints on valid event graphs. These constraints must be respected by valid NuHepMC events as full compatibility with HepMC3 is required. More details of these constraints can be found in Ref.~\cite{Buckley2021}.

Existing neutrino event generators often rely on effective descriptions of the nuclear environment in a neutrino--nucleus hard scattering process. This means that four-momentum may not be explicitly conserved for the neutrino--nucleus system. Energy and momentum can be exchanged with a \emph{nuclear remnant}, which is often not treated carefully in the simulation of a primary neutrino--nucleon collision, via initial- and final-state interactions. Implementations are of course free to thoroughly model all involved particles, conserving four-momentum over the whole system, including a fully-simulated nuclear remnant. For those implementations where such a requirement is not feasible or would delay the adoption of this standard, \ref{part:nuclblob} reserves a non-standard particle number that can be used to represent a nuclear remnant that is not precisely simulated.

% E.R.2
\item \textsc{Event Number}:\\ \label{evt:num}\noindent
Each \texttt{HepMC3::GenEvent} must have a non-negative event number
that is unique within a given vector.

% E.R.3
\item \textsc{Process ID}:\\ \label{evt:proc_id}\noindent
The process ID for the primary physics process that is represented in the
\\ \texttt{HepMC3::GenEvent} must be recorded in a \texttt{HepMC3::IntAttribute}
named \texttt{"signal\_process\_id"}.
%~\footnote{This attribute is similar to that used in the LHC community. Hence, the use of snake case instead of camel case like the other attributes in the standard.}
The metadata describing this process ID must be stored according to~\ref{gen:proc_id}.

% E.R.4
\item \textsc{Units}:\\ \label{evt:units}\noindent
Energy and position units must be explicitly set in the \texttt{HepMC3::GenEvent}.

% E.R.5
\item \textsc{Lab Position}:\\ \label{evt:lab_pos}\noindent
The position of the event in the lab frame must be added as a
\\ \texttt{HepMC3::VectorDoubleAttribute}, named \texttt{"lab\_pos"}, with the same units as used when implementing~\ref{evt:units}. See~\ref{evt:lab_time} for how to optionally store time in this attribute. If the simulation did not involve a macroscopic geometry, then this variable may be set to \texttt{[0, 0, 0]}.

% E.R.6
\item \textsc{Vertices}:\\ \label{evt:vert}\noindent
An event must contain at least one \texttt{HepMC3::GenVertex}, and must have
one and only one with a \emph{primary interaction vertex} status code. No \texttt{HepMC3::GenVertex} may have a \emph{not defined} status code. (See~\ref{vert:status} for additional details).

% E.R.7
\item \textsc{Beam and Target Particles}:\\ \label{evt:beam_target}\noindent
An event must contain exactly one particle with the \emph{incoming beam
particle} status code and one particle with the \emph{target particle} status
code (see~\ref{part:status}). It is recommended that, in cases where the colliding
initial-state particles are distinct, the more massive of the two should be
considered the target. For neutrino scattering, the target will thus often be a
complex nucleus or a free nucleon. In the case of equally massive particles,
the choice to label one of them as the target is arbitrary.

\textit{n.b.} A nucleon (or quark) bound within a nucleus should never be marked as the target particle; the nucleus itself should be considered the target. \ref{part:struck_nuc} provides a convention for marking a constituent bound
nucleon struck by the incoming beam particle in the event graph.

% E.R.8
\item \textsc{Event Completeness}:\\ \label{evt:completeness}\noindent
All simulated incoming and outgoing physical particles must be written to the
event. The storage of intermediate particles is considered an implementation
detail.

% E.R.9
\item \textsc{Event Frame of Reference}:\\ \label{evt:frame}\noindent
As the primary use of HepMC3 is for experiment simulation, all particle momenta specified via standard HepMC3 mechanisms must be defined in the lab frame. From the information required by a standard HepMC3 event, arbitrary frame transformations should be possible and can be performed by the \\ \texttt{HepMC3::GenEvent::boost} method. If storing kinematic information in other frames of reference facilitates later calculations, this should be done via simulation-specific \texttt{HepMC3::Attribute}s on the relevant
\texttt{HepMC3::GenParticle}s.

\end{enumerate}

\subsubsection{Conventions}
\begin{enumerate}[label={\textsc{E.C.\arabic*}},leftmargin=3.5\parindent]

% E.C.1
\item \textsc{Process IDs}:\\ \label{evt:proc_id_convention}\noindent
It is not appropriate to mandate a specific set of interaction processes and
assign them IDs in this standard. Different models make different choices, and
it is impossible to foresee modeling developments that would require new process
IDs to be defined in the future. Instead, the ranges of IDs given below are provided as a categorization of processes to facilitate high-level analysis. If a given process naturally fits in one of these categories it can be helpful to choose IDs that follow this convention, but an explicit aim of this standard is to avoid constraining or defining what processes can be stored and described. \textit{n.b.} Even if an
implementation uses the convention in Table~\ref{tab:proc_id}, it must still
adhere to~\ref{gen:proc_id}.

\begin{table}[ht]
\centering
    \begin{tabular}{|c | c|}
        \hline
        \textbf{Identifier} & \textbf{Process} \\
        \hline
        100-199 & Low-energy nuclear scattering \\
        200-299 & Quasi-elastic \\
        300-399 & Multi-body scattering \\
        400-499 & Baryon resonance production \\
        500-599 & Shallow inelastic scattering \\
        600-699 & Deep inelastic scattering \\
        700-    & Other process types \\
        \hline
    \end{tabular}
    \caption{Optional set of identifier ranges for various high-level process categories.}
    \label{tab:proc_id}
\end{table}

If following this convention, charged-current (CC) processes should have identifiers in the X00-X49 block
and neutral-current (NC) processes should have them in the X50-X99 block.
Negative process IDs should be used for electromagnetic interactions in
event generators that include them.

In Table~\ref{tab:proc_id},
distinct ranges for shallow inelastic scattering and deep inelastic scattering are provided because existing neutrino generators often make such a distinction in both the modeling details and simulated event metadata. It should be noted that the boundary between these two is not \textit{a priori} well defined and choices varies significantly in current neutrino scattering simulations, usually defined with reference to some boundary value in the invariant mass of the hadronic final state. The lack of consensus and common definitions hinders progress in modeling neutrino-induced low energy hadronization.
New efforts in this area are sorely needed and would facilitate collaboration with the collider community, as discussed in Section 2.4 of Ref.~\cite{Campbell:2022qmc}.

% E.C.2
\item \textsc{Total Cross Section}:\\ \label{evt:total_xsec}\noindent
The total cross section for the incoming beam particle, with its specific energy, to interact with the
target particle should be stored in a \texttt{HepMC3::DoubleAttribute} on the
\texttt{HepMC3::GenEvent}, named \texttt{"tot\_xs"}. See \ref{gen:cs_units} for how to communicate cross section units.

% E.C.3
\item \textsc{Process Cross Section}:\\ \label{evt:proc_xsec}\noindent
The total cross-section for the selected process ID for the incoming beam
particle, with its specific energy, to interact with the target particle should be stored in a
\texttt{HepMC3::DoubleAttribute} on the \texttt{HepMC3::GenEvent}, named
\texttt{"proc\_xs"}. See \ref{gen:cs_units} for how to communicate cross section units.

% E.C.4
\item \textsc{Estimated Flux-Averaged Total Cross Section}:\\
\label{evt:est_xsec}\noindent
Some simulations build up an estimate of the flux-averaged total cross section
\AverageXSec\ as they run, which makes implementing~\ref{gen:fatcs} impractical
in many cases. As an alternative, the built-in attribute
\texttt{HepMC3::GenCrossSection}, accessed via
\texttt{GenEvent::cross\_section}, should be used to store the current estimate
of \AverageXSec. A user should only ever need to use the estimate provided with
the last event that they read to correctly scale an event rate to a cross-section prediction.
This means that statistically correct predictions can be made without reading an event vector to
the end. The \\\texttt{HepMC3::GenCrossSection::cross\_section\_errors} data member can be used to
decide when enough events have been read to reach some desired statistical precision
on the total cross section. The best estimate from the generator run will always be
provided on the final event in a vector.

For event generators that do not currently provide the value of \AverageXSec\ in the output, Appendix~\ref{app:av_xsec_mc} provides an algorithm for computing a running estimate, with associated Monte Carlo statistical uncertainty,
as events are generated.

When implementing this convention, ensure that the \texttt{cross\_sections} and
\\\texttt{cross\_section\_errors} data members are the same length as the number
of weights defined in the header. These should be filled with the current
estimate of the total cross section for each variation based on all events
generated so far, including the current event. Additionally, the
\texttt{HepMC3::GenCrossSection} data members \texttt{accepted\_events} and
\texttt{attempted\_events} should be filled with appropriate values.

% E.C.5
\item \textsc{Lab Time}:\\ \label{evt:lab_time}\noindent
If the \texttt{"lab\_pos"} attribute vector contains three entries then it is assumed to only contain the spatial position of the event. If it contains four entries, then the fourth entry is interpreted as the time of the event in seconds.

\end{enumerate}

\subsubsection{Suggestions}
\begin{enumerate}[label={\textsc{E.S.\arabic*}},leftmargin=3.5\parindent]

% E.S.1
\item \textsc{Beam Description (Beam Simulation)}\\ \label{evt:beam_description}\noindent
For more complex beam simulations that can not adequately be described by a single energy or energy histogram (see \ref{gen:beam_energy}), it is suggested that the full parent decay history is included in the \texttt{HepMC3::GenEvent}. A full set of conventions for the description of beam particle production and parent particle decay chains (for the case of neutrino beams) is currently outside the scope of this specification, but generator implementations can signal that they adhere to this suggestion to notify users that some or all of the beam particle production information is included in the event.

\end{enumerate}

\subsection{Vertex Information}
\label{subsec:vert}

The vertices in a HepMC3 event are used to connect groups of incoming and
outgoing particles. For the vertex information, there is only one requirement and one convention
in the present version of the NuHepMC standard.

\subsubsection{Requirements}
\begin{enumerate}[label={\textsc{V.R.\arabic*}},leftmargin=3.5\parindent]

% V.R.1
\item \textsc{Vertex Status Codes}:\\ \label{vert:status}\noindent
The standard HepMC3 range of reserved values for \texttt{HepMC3::GenVertex::status} is extended to
include the concept of a primary vertex, corresponding to the \emph{primary}
hard-scattering process (\textit{i.e.}, the one labeled according to \ref{evt:proc_id}), and a final state interaction (FSI) summary vertex. The full set of
defined status codes can be found in Table~\ref{tab:vert_status}.
Implementations are free to define specific vertex status codes to refer to
individual FSI (or ISI) processes and output as much information as they
require. However, a single summary vertex may be useful for some purposes if the full FSI history is very detailed or not often needed by users. See Figure \ref{fig:neutevg} for an example of an event with such a summary vertex.

\begin{table}[ht]
    \centering
    \begin{tabular}{|c|c|c|}
        \hline
        Status Code & Meaning & Usage  \\
        \hline
        0 & Not defined & Do not use \\
        1 & Primary interaction vertex & Recommended for all cases \\
        2 & FSI Summary vertex & Recommended for all cases \\
        3-20 & Reserved for future NuHepMC standards & Do not use \\
        21-999 & Generator-dependent & For generator usage \\
        \hline
    \end{tabular}
    \caption{Set of vertex status codes}
    \label{tab:vert_status}
\end{table}

Any secondary vertex included within a NuHepMC event may have a status between
21 and 999. \ref{gen:vert_status} requires that all
generator-specific status codes must be fully described by attributes stored
in the \texttt{HepMC3::GenRunInfo}.

\end{enumerate}

\subsubsection{Conventions}
\begin{enumerate}[label={\textsc{V.C.\arabic*}},leftmargin=3.5\parindent]

% V.C.1
\item \textsc{Bound Nucleon Separation Vertex}\\ \label{vert:iavertex}\noindent
When an interaction with a nucleon bound within a nucleus with definite kinematics is simulated, a \texttt{HepMC3::GenVertex} corresponding to the separation of the struck nucleon and the nuclear remnant may be included and assigned status code \num{21}. If this convention is signaled via \ref{gen:convention}, then status code \num{21} need not be included in the implementation of \ref{gen:vert_status}.

\end{enumerate}

\subsection{Particle Information}
\label{subsec:part}

In the current version of the NuHepMC standard, there is only a single
requirement and two conventions for the particle information.

\subsubsection{Requirements}
\begin{enumerate}[label={\textsc{P.R.\arabic*}},leftmargin=3.5\parindent]

% P.R.1
\item \textsc{Particle Status Codes}:\\ \label{part:status}\noindent
The standard HepMC3 range of reserved values for \\ \texttt{HepMC3::GenParticle::status} is
slightly extended to include the concept of a target particle. For neutrino scattering,
this will usually be a target nucleus. The status codes are defined in
Table~\ref{tab:part_status}.

\begin{table}[ht]
    \centering
    \begin{tabular}{|c|c|c|}
        \hline
        Status Code & Description & Usage  \\
        \hline
        0 & Not defined & Do not use \\
        1 & Undecayed physical particle & Recommended for all cases \\
        2 & Decayed physical particle & Recommended for all cases \\
        3 & Documentation line & Used for in/out particles  \\
          &                    & in the primary process \\
        4 & Incoming beam particle & Recommended for all cases \\
        5-10 & Reserved for future HepMC3 standards & Do not use \\
        11-19 & Reserved for future NuHepMC standards & Do not use \\
        20 & Target particle & Recommended for all cases \\
        21-200 & Generator-dependent & For generator usage \\
        201- & Simulation-dependent & For simulation software usage \\
        \hline
    \end{tabular}
    \caption{Particle status codes}
    \label{tab:part_status}
\end{table}

Note especially that any incoming real particle must have a status code of 4 or
20, and any outgoing real particle must have a status code of 1. Special
care must be taken when including the effects of initial-state and final-state
interactions.

Any internal particle included within a NuHepMC event may have a status in the range
than 21-200. \ref{gen:part_status} requires that all generator-specific
status codes must be fully described by attributes stored in the
\texttt{HepMC3::GenRunInfo}.

\end{enumerate}

\subsubsection{Conventions}
\begin{enumerate}[label={\textsc{P.C.\arabic*}},leftmargin=3.5\parindent]

% P.C.1
\item \textsc{Particle Status Codes}:\\ \label{part:struck_nuc}\noindent
When an interaction with a bound nucleon with definite kinematics is simulated,
the internal \texttt{HepMC3::GenParticle} corresponding to the bound nucleon
should have status code 21. If this convention is signaled via \ref{gen:convention}, then status code 21 need not be included in
the implementation of \ref{gen:part_status}.

% P.C.2
\item \textsc{Nuclear Remnant Particle Code}:\\ \label{part:nuclblob}\noindent
HepMC3 places restrictions on all external particles in the event graph to facilitate automatic checking of four momentum conservation at the event graph level. As a result, a new particle number 2009900000 is defined to identify a nuclear remnant that is not precisely handled by the interaction simulation. If a generator does correctly simulate any nuclear remnant, such that the total mass and energy are well defined, then this convention is unnecessary and the remnant can be written out as a particle with a standard PDG nuclear number and mass and momentum set as normal. A pseudoparticle with this number is explicitly an implementation detail that should be used to abide by HepMC3 constraints to not have vertices with no outgoing particles. Any pseudoparticle with this number should not be considered for physics analyses or onward simulation. The number is chosen, according to the PDG scheme~\cite[Sec.~45, p.~733]{PDG}, to be outside the range reserved for nuclear and quark-content particles. Its status as a non-standard particle code is signaled by the 6th and 7th least significant digits being set to 9.

If the the kinematics of the nuclear remnant are not known, but the proton and neutron numbers are well-defined, then the corresponding PDG nuclear number may be added as a \texttt{HepMC3::IntAttribute} on the \texttt{HepMC3::GenParticle} named \texttt{"remnant\_particle\_number"}.

If this convention is signaled via the mechanism
described in \ref{gen:convention}, then particle number 2009900000 need not be included in
the implementation of \ref{gen:pdg_nums}.

\end{enumerate}

%\section{Implementation Example}
%\label{sec:impl}
%
%\section{Validator Tool}
%\label{sec:valid}
%
%We provide a tool to validate that the HepMC3 event vector satitisfies the NuHepMC3 standard described in this work at \url{https://github.com/NuHepMC/ReferenceImplementation/blob/main/NuHepMCReferenceValidator.cxx}.
%
%\section{Usage in Tools}
%
%\subsection{GENIE}
%
%\subsection{NuWro}
%
%\subsection{NEUT}
%
%\subsection{Achilles}
%
%\subsection{Nuisance}

\section{Flux-averaged total cross section}
\label{sec:flux_averaged_xsec}

Comparisons of event generator predictions to external model calculations and
experimental data typically involve the conversion of simulated event
distributions to total or differential cross sections. This conversion is
usually made using a scaling factor \AverageXSec\ called the flux-averaged
total cross section. This factor is simple to use but often difficult to
calculate analytically. Given the importance of \AverageXSec\ for analyses of
simulated neutrino scattering events, we provide mathematical details about its
definition and calculation.

Section~\ref{sec:av_xsec_def} derives an expression for \AverageXSec\ that is
generally applicable to most simulations of interest for neutrino experiments,
including those with a time-dependent neutrino source and a full treatment of
the detector geometry. Example methods for obtaining Monte Carlo estimators of \AverageXSec\ suitable for storage via \ref{evt:est_xsec} are described in Appendix~\ref{app:av_xsec_mc}. Section~\ref{sec:av_xsec_analytic} describes some simple cases in which
\AverageXSec\ may be calculated directly, making them good candidates for
situations in which event generators may implement~\ref{gen:fatcs}.

\subsection{Derivation}
\label{sec:av_xsec_def}

Let $\DiffFlux(\NuSpecies, \NuEnergy, \NuTheta, \NuPhi, \NuPosition, \NuTime)$
be the differential flux of (anti)neutrinos of species \NuSpecies\ with energy
\NuEnergy, momentum direction (expressed in terms of the Cartesian unit
vectors)
\begin{equation}
\NuDirection = \sin\NuTheta \cos\NuPhi \, \mathbf{\hat{x}}
+ \sin\NuTheta \sin\NuPhi \, \mathbf{\hat{y}}
+ \cos\NuTheta \, \mathbf{\hat{z}}\,,
\end{equation}
and instantaneous three-position \NuPosition\ at time \NuTime. This quantity is
defined so that
\begin{equation}
\IntegratedFlux(\NuPosition) = \sum_\NuSpecies \int d\NuEnergy
\, d\cos\NuTheta \, d\NuPhi \, d\NuTime \, \DiffFlux
\end{equation}
has units of integrated flux (\textit{e.g.}, \si{\per\centi\meter\squared}).

Assuming that the relevant neutrino interaction cross sections are sufficiently
small that beam attenuation and multiple scattering effects can be neglected,
then the total number of interactions \NumInteractions\ expected in a
volume of interest
\begin{equation}
V = \int d^3\NuPosition
\end{equation}
is given by
\begin{equation}
\label{eq:NumInteractions1}
\NumInteractions = \sum_\NuSpecies \sum_\TargetSpecies \int d\NuEnergy \,
d\cos\NuTheta \, d\NuPhi \, d^3\NuPosition \, d\NuTime \,
\DiffFlux \, \TargetDensity \, \TotalXSec\,\,
\end{equation}
where $\TargetDensity = \TargetDensity(j,\NuPosition)$ is the number density of
the \TargetSpecies-th kind of target at position \NuPosition. The symbol
$\TotalXSec = \TotalXSec(\NuSpecies,\TargetSpecies,\NuEnergy)$ denotes the
total cross section for (anti)neutrinos of species \NuSpecies\ and energy
\NuEnergy\ to interact with the \TargetSpecies-th kind of target (typically a
particular nuclide).

One may rewrite the expression for \NumInteractions\ in the simple form
\begin{equation}
\NumInteractions = \AverageFlux \cdot \AverageTargets \cdot \AverageXSec
\end{equation}
where
\begin{equation}
\label{eq:AvFlux}
\AverageFlux = \frac{1}{V}\sum_\NuSpecies \int d\NuEnergy \,
d\cos\NuTheta \, d\NuPhi \, d^3\NuPosition \, d\NuTime \, \DiffFlux
\end{equation}
is the average flux in the volume $V$,
\begin{equation}
\label{eq:AvTargets}
\AverageTargets = \frac{1}{\AverageFlux} \sum_\NuSpecies \sum_\TargetSpecies
\int d\NuEnergy \, d\cos\NuTheta \, d\NuPhi \, d^3\NuPosition \,
d\NuTime \, \DiffFlux \, \TargetDensity
\end{equation}
is the flux-averaged number of targets illuminated by the (anti)neutrinos,
and
\begin{equation}
\label{eq:AvXSec}
\AverageXSec = \frac{\NumInteractions}{\AverageFlux \cdot \AverageTargets}
\end{equation}
is the flux-averaged total cross section. For a discussion on how these might
be calculated using Monte-Carlo methods see Appendix~\ref{app:av_xsec_mc}.

\subsection{Analytic calculation under simplifying assumptions}
\label{sec:av_xsec_analytic}

In certain simple cases, the flux-averaged total cross section \AverageXSec\ is
known at the start of the run and can be computed analytically to implement
\ref{gen:fatcs}. A common example is the case of a point target of type
$b$ that is exposed to uni-directional beam of (anti)neutrinos of
species $a$ with a fixed emission time. In this case, we may
write the differential flux and target density as
\begin{align}
\DiffFlux &= A(\NuEnergy) \, \delta(\cos\NuTheta - \cos\theta_0)
\, \delta(\NuPhi - \phi_0) \, \delta(\NuTime - t_0)
\, \delta_{fa} \\
\TargetDensity &= B \, \delta^3(\NuPosition - \NuPosition_0) \, \delta_{jb} \,.
\end{align}
Here $\delta_{xy}$ is the Kronecker delta and the constants $\theta_0$,
$\phi_0$, $t_0$, $\NuPosition_0$, $a$, and $b$ are arbitrary values of the
relevant variables. The function $A(\NuEnergy)$ gives the incident
(anti)neutrino energy spectrum (with unimportant normalization) and has units
of flux divided by energy (\textit{e.g.}, \si{\per\GeV\per\centi\meter\squared}). The
constant $B$ is dimensionless. From Eqs.~\ref{eq:AvFlux}, \ref{eq:AvTargets},
and \ref{eq:AvXSec}, it follows that
\begin{align}
\AverageFlux &= \int d\NuEnergy \, A(\NuEnergy) \\[2mm]
\AverageTargets &= B \\[2mm]
\label{eq:average_xsec_analytic}
\AverageXSec &= \frac{\int d\NuEnergy \, A(\NuEnergy)
\, \TotalXSec(a,b,\NuEnergy)}
{\int d\NuEnergy \, A(\NuEnergy)} \,.
\end{align}
If the function $A(\NuEnergy)$ and the total cross section
$\TotalXSec(a,b,\NuEnergy)$ are accessible during generator initialization,
then the integrals in Eq.~\ref{eq:average_xsec_analytic} are easily computable
by standard numerical methods.

In the even simpler case where the neutrino source is also monoenergetic, then
$A(\NuEnergy)$ contains a Dirac delta function, and the flux-averaged total
cross section is just the total cross section evaluated at the fixed energy of
the beam.

\section{NuHepMC Generator Predictions}
\label{sec:nuiscomp}

Using preliminary tools for converting proprietary neutrino event formats to the NuHepMC standard defined above, we demonstrate the utility of the common format with some cross-section predictions.
See Tab.~\ref{tab:support} for the current status of implementations of NuHepMC in the most widely-used neutrino event generators.~\footnote{An up to date table of the current status of adoption within the generators can be found in the NuHepMC Spec repository: \url{https://github.com/NuHepMC/Spec}.}
It is important to note that once built-in support for NuHepMC is adopted within a generator, the public converter from the previous generator-specific formats to NuHepMC will be deprecated.

\begin{table}[ht]
    \centering
    \begin{tabular}{|c|c|c|}
    \hline
     Generator    & Output Formats & Public Converter  \\
     \hline
     Achilles    & NuHepMC & N/A \\
     GENIE & GHep (ROOT \texttt{TTree}), NuHepMC in review & No \\
     GiBUU & LHA XML, NuHepMC in 2025 release & N/A \\
     MARLEY & NuHepMC (in upcoming v2 release) & N/A \\
     NEUT & \texttt{neutvect} (ROOT \texttt{TTree}) & Yes \\
     NuWro & \texttt{event1} (ROOT \texttt{TTree}) & Yes \\
    \hline
    \end{tabular}
    \caption{Current state of output formats and public converters available in neutrino event generators. Public converters are from default formats to NuHepMC.}
    \label{tab:support}
\end{table}

\begin{figure}
\centering
    \includegraphics[width=0.49\textwidth]{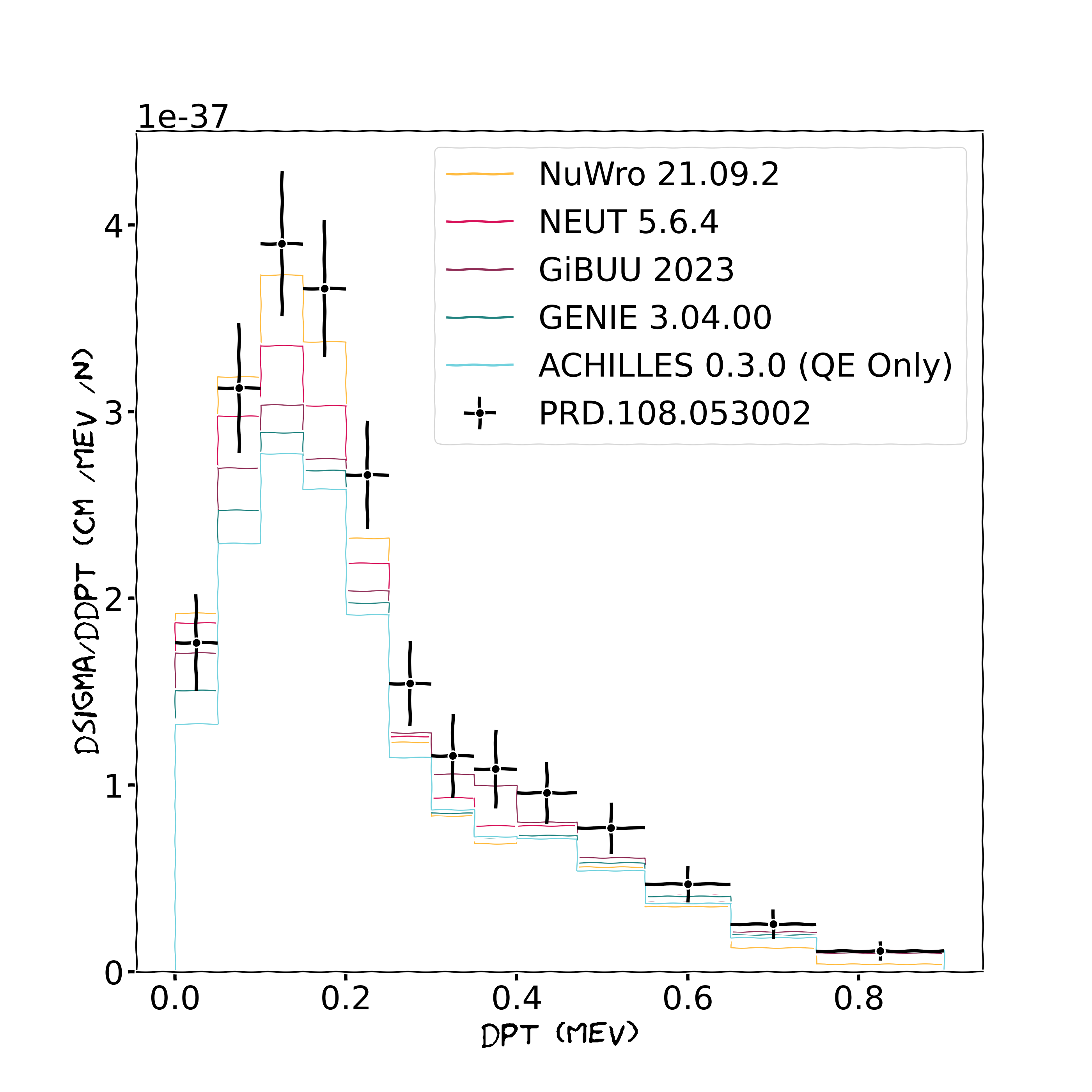}
    \includegraphics[width=0.49\textwidth]{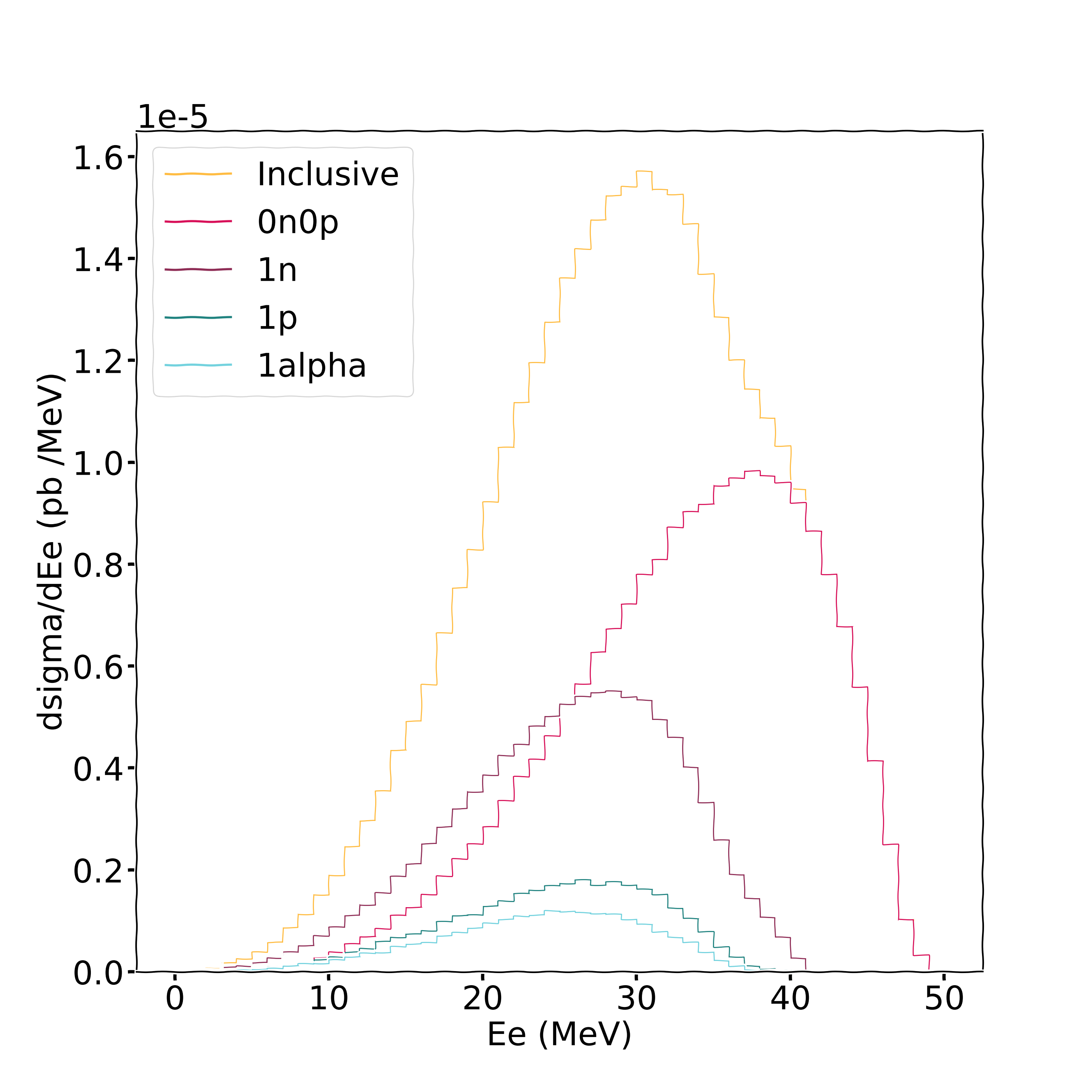}
    \caption{(\emph{left}) Comparison of NuWro, NEUT, GiBUU, GENIE, and ACHILLES to the $\delta p_T$ data from the \mbox{MicroBooNE} collaboration~\cite{PhysRevD.108.053002}. The comparison is a demonstration of the event generation to NUISANCE~\cite{Stowell:2016jfr} pipeline and not a physics statement about the prediction quality of any generator. (\emph{right}) Low-energy differential cross section predictions produced using MARLEY events in the NuHepMC format.\protect\footnotemark}
    \label{fig:predictions}
\end{figure}

A comparison to a recent neutrino-argon cross-section measurement from the \mbox{MicroBooNE} collaboration~\cite{PhysRevD.108.053002} is shown on the left-hand side of Figure~\ref{fig:predictions}. This comparison was made with the NUISANCE~\cite{Stowell2017} framework,\footnote{The NUISANCE framework can be obtained at: \url{https://github.com/NUISANCEMC/nuisance}.} which before this implementation of NuHepMC would have to have been built against GENIE, NEUT, and NuWro binaries of compatible versions to be able to generate the predictions shown in the figure. In the present workflow, results can be obtained using only generator-agnostic tools for interpreting events stored in the NuHepMC format. This reduction of technical requirements will dramatically lower the barrier to making high-quality prediction--data comparisons for the neutrino community.

It is particularly worth noting that the different generators included in this comparison provide critical cross-section scaling information in different ways. Thanks to the provisions in this standard, NUISANCE is able to use \ref{gen:convention} to determine which representation of \AverageXSec\ is available and then automatically scale the selected event rate to a differential cross-section prediction without any user input or generator-specific code or knowledge. For reference, the NEUT and NuWro event distributions are scaled via \ref{gen:fatcs}, while GiBUU, GENIE, and Achilles are scaled via \ref{evt:est_xsec}.

We want to emphasize that the comparison to data is solely a demonstration of the prediction pipeline and not meant as a physics statement on the best or most recent predictions from each of the generators for the given measurement.

The right-hand side of Figure~\ref{fig:predictions} presents MARLEY predictions of flux-averaged differential cross sections for $\nu_e$ produced from $\mu^{+}$ decay at rest scattering on \isotope[40]{Ar}.~\footnotetext{The analysis for \mbox{MicroBooNE} is in the NUISANCE analysis framework. The analysis for MARLEY can be found at~\url{https://github.com/NuHepMC/cpputils/blob/main/examples/Marley.ipynb}} The yellow histogram shows the inclusive charged-current prediction, and the other histograms show contributions arising from several distinct final states. The cross-section predictions obtained using NuHepMC without any MARLEY-specific code are very similar to those shown in a previous MARLEY publication~\cite{marleyPRC}.

Comparisons similar to those above can be easily produced for an arbitrary interaction channel using information available in the NuHepMC event record. A Monte Carlo estimator for the flux-averaged differential cross
section in a bin of an arbitrary kinematic variable $x \in [x_k, x_{k+1})$
is given by
\begin{equation}
\left< \frac{d\sigma}{dx} \right>_{\!k}
\approx \frac{ \AverageXSec \, \sum_{\EventIndex=1}^{\NumEvents}
\delta^k_\EventIndex \, w_\EventIndex }{ \Delta x_k \, \sum_{\EventIndex=1}^{\NumEvents} w_e  }
\end{equation}
where \AverageXSec\ is the flux-averaged total cross section (see
Sec.~\ref{sec:flux_averaged_xsec}), $w_\EventIndex$ is the statistical weight
of the \EventIndex-th event, \NumEvents\ is the number of generated events, and $\Delta x_k = x_{k+1} - x_k$ is the width of the $k$-th bin. The symbol $\delta^k_e$ evaluates to unity when the value of $x$ from the $e$-th
event falls within the $k$-th bin and zero otherwise.
In the case of unit-weight events, the value of $w_e = 1$ for all events, and $\sum_{\EventIndex=1}^{\NumEvents} w_e = n$.
One thus recovers the traditional definition of these estimators for unweighted events.

The Monte Carlo
statistical uncertainty (standard deviation) on the estimator of the differential cross section is approximately
given by
\begin{equation}
\text{StdDev}\!\left(
\left< \frac{d\sigma}{dx} \right>_{\!k}\right)
\approx \frac{ \AverageXSec }{ \Delta x_k \, \sum_{\EventIndex=1}^{\NumEvents} w_\EventIndex}
\, \sqrt{ \sum_{\EventIndex=1}^{\NumEvents} \delta^k_\EventIndex
\, w^2_\EventIndex }
\end{equation}
where we have assumed that \AverageXSec\ is exactly known. For cases where it
is estimated using Monte Carlo techniques rather than directly calculated, the
statistical uncertainty discussed in Appendix~\ref{app:av_xsec_mc} also applies.

\section*{Acknowledgements}
The authors would like to thank A. Papadopoulou for sharing a NUISANCE
implementation for the example comparison in Sec.~\ref{sec:nuiscomp}; A. Verbytskyi for reading and commenting on this manuscript; and J. Tena Vidal for implementing an early version of this standard to test its practicality for a non-author. This
manuscript has been authored by Fermi Research Alliance, LLC under Contract No.
DE-AC02-07CH11359 with the U.S. Department of Energy, Office of Science, Office
of High Energy Physics. The work in this manuscript is supported by the Royal
Society, grant number URF{\textbackslash}R1{\textbackslash}211661.

\begin{appendix}
\numberwithin{equation}{section}

\section{Estimation via Monte Carlo sampling}
\label{app:av_xsec_mc}

In this section we provide two example methods of estimating the flux-averaged total cross-section when it is not known \emph{a priori}. The first method is suitable for  GENIE-like codes that calculate energy-dependent total cross sections prior to event generation itself. The second is suitable for ACHILLES-like codes that do not use pre-calculated total cross sections in this way.

\subsection{GENIE-like approach}

The (anti)neutrino interactions that occur in the volume $V$ are drawn from the
probability distribution
\begin{equation}
\ProbDist = \frac{1}{\NumInteractions} \, \DiffFlux \, \TargetDensity
\, \TotalXSec\,.
\end{equation}
It follows from the definitions in Sec.~\ref{sec:flux_averaged_xsec} that
\begin{equation}
\AverageXSec = \Bigg[ \sum_\NuSpecies \sum_\TargetSpecies \int \ProbDist \,
\frac{1}{\TotalXSec} \, d\NuEnergy \, d\cos\NuTheta \, d\NuPhi \,
d^3\NuPosition \, d\NuTime \Bigg]^{-1} \,.
\end{equation}
One may therefore obtain a Monte Carlo estimator \AverageXSecEstimator\ for the
flux-averaged total cross-section \AverageXSec\ via the expression
\begin{equation}
\label{eq:fatxs_MC}
\AverageXSecEstimator = \left[ \frac{
\sum_{\EventIndex=1}^{\NumEvents} \frac{w_\EventIndex}
{\TotalXSec_\EventIndex(\NuSpecies,
\TargetSpecies,\NuEnergy)} }{\sum_{\EventIndex=1}^{\NumEvents} w_\EventIndex}
\right]^{-1}
\end{equation}
where $w_\EventIndex$ is the statistical weight for the $\EventIndex$-th event,
\NumEvents\ is the total number of generated events, and
$\TotalXSec_\EventIndex$ is the total cross section evaluated for the
(anti)neutrino species (\NuSpecies), target (\TargetSpecies), and incident
energy (\NuEnergy) sampled in the \EventIndex-th event. An estimator for the
Monte Carlo statistical standard deviation of \AverageXSecEstimator\ is given
by
\begin{equation}
\label{eq:fatxs_MC_stddev}
\text{StdDev}(\AverageXSecEstimator) =
\sqrt{ \text{Var}(\AverageXSecEstimator) }
= \AverageXSecEstimator^2 \cdot
\sqrt{ \text{Var}\!\left(\frac{1}{\AverageXSecEstimator}\right) }
= \AverageXSecEstimator^2 \cdot \sqrt{ \frac{n \cdot
\sum_{\EventIndex=1}^{\NumEvents} w_\EventIndex^2
\left(\frac{1}{\TotalXSec_\EventIndex}
- \frac{1}{\AverageXSecEstimator}\right)^2}
{\left( \sum_{\EventIndex=1}^{\NumEvents} w_\EventIndex
\right)^2 \cdot (n - 1)} } \,.
\end{equation}

%\subsection{Choice of standard error estimator}
%\label{sec:av_xsec_error_note}

Note that there is no universally accepted expression for the standard error on
a weighted arithmetic mean such as the one computed in Eq.~\ref{eq:fatxs_MC}
(before taking the reciprocal of the expression). The choice used in
Eq.~\ref{eq:fatxs_MC_stddev} is based on the third expression for
``$\mathrm{SEM_w}$'' recommended in Ref.~\cite{Gatz1995} based on bootstrapping
studies (see also references therein). The quantity proposed as an estimator
for the square of the standard error on a weighted mean
\begin{equation}
\bar{x}_w = \frac{\sum_{\EventIndex=1}^\NumEvents w_\EventIndex
\, x_\EventIndex }{\sum_{\EventIndex=1}^\NumEvents w_\EventIndex }
\end{equation}
of $\NumEvents$ values $x_\EventIndex$ is
\begin{equation}
\label{eq:StatVar}
\text{Var}({\bar{x}_w}) = \frac{\NumEvents}{(\NumEvents - 1)
\left( n \bar{w} \right)^2 }
\left[\sum_{\EventIndex=1}^{\NumEvents} \left(w_\EventIndex x_\EventIndex
- \bar{w} \bar{x}_w\right)^2
- 2\bar{x}_w(w_\EventIndex - \bar{w})(w_\EventIndex x_\EventIndex
- \bar{w} \bar{x}_w) + \bar{x}_w^2(w_\EventIndex - \bar{w})^2
\right]
\end{equation}
where
\begin{equation}
\bar{w} = \frac{1}{n} \sum_{\EventIndex=1}^{\NumEvents} w_\EventIndex \,.
\end{equation}
The expression in Eq.~\ref{eq:StatVar} may be simplified to read
\begin{equation}
\text{Var}({\bar{x}_w}) = \frac{\NumEvents}{(\NumEvents - 1)
\left( \sum_{\EventIndex=1}^{\NumEvents} w_\EventIndex \right)^2 }
\sum_{\EventIndex=1}^{\NumEvents} w_\EventIndex^2 \,
(x_\EventIndex - \bar{x}_w)^2 \,.
\end{equation}
The result in Eq.~\ref{eq:fatxs_MC_stddev} is obtained immediately with the
substitutions $x_\EventIndex \to 1/\TotalXSec_\EventIndex$ and $\bar{x}_w \to
1/\AverageXSecEstimator$.

%\subsection{Algorithm for running estimation}
%\label{sec:av_xsec_run}

While the expressions in Eqs.~\ref{eq:fatxs_MC} and \ref{eq:fatxs_MC_stddev}
may be useful for estimating \AverageXSec\ from a sample of generated events,
they require access to the entire sample and are thus unsuitable for
implementing the running estimate described in~\ref{evt:est_xsec}. However, an
adaptation of West's algorithm~\cite{West1979} provides a solution for this
application. Let
\begin{equation}
n_0 = S_0 = \mu_0 = T_0 = 0\,.
\end{equation}
Then, for the \EventIndex-th event, let the values of these quantities
be defined recursively via
\begin{align}
n_\EventIndex &= n_{\EventIndex - 1} + 1 = \EventIndex \\
S_\EventIndex &= S_{\EventIndex - 1} + w_\EventIndex \\
\mu_\EventIndex &= \mu_{\EventIndex - 1} + \frac{ w_\EventIndex }
{ S_\EventIndex } \,
\left(\frac{1}{\TotalXSec_\EventIndex} - \mu_{\EventIndex - 1}\right) \\[1mm]
T_\EventIndex &= T_{\EventIndex - 1} + w_\EventIndex^2 \,
\left(\frac{1}{\TotalXSec_\EventIndex} - \mu_{\EventIndex - 1}\right)
\left(\frac{1}{\TotalXSec_\EventIndex} - \mu_{\EventIndex}\right) \,,
\end{align}
where $\TotalXSec_\EventIndex$ and $w_\EventIndex$ are assigned the same
meanings as above.

The running Monte Carlo estimator $\AverageXSecEstimator_\EventIndex$ of
\AverageXSec\ for the \EventIndex-th event ($\EventIndex > 0$) may then be
written as
\begin{equation}
\AverageXSecEstimator_\EventIndex = \frac{1}{\mu_\EventIndex} \,.
\end{equation}
Its estimated statistical uncertainty is given by
\begin{equation}
\text{StdDev}(\AverageXSecEstimator_\EventIndex) =
\frac{1}{ \mu_\EventIndex^2 } \,
\sqrt{ \frac{n_\EventIndex \, T_\EventIndex }{(n_\EventIndex-1) \,
S_\EventIndex^2 }} \,.
\end{equation}

\subsection{ACHILLES-like approach}

When calculating Eq.~\eqref{eq:NumInteractions1}, the equation can be expanded to include the
integrals over the differential cross section giving
\begin{equation}
\NumInteractions = \sum_\NuSpecies \sum_\TargetSpecies \int d\NuEnergy \,
d\cos\NuTheta \, d\NuPhi \, d^3\NuPosition \, d\NuTime d\Omega \,
\DiffFlux_\NuSpecies \, \TargetDensity_\TargetSpecies \, \frac{d\sigma_{\NuSpecies\TargetSpecies}}{d\Omega} \,,
\end{equation}
where $\Omega$ is the final state phase space and the cross section is broken up into individual
processes as $\sigma_{\NuSpecies\TargetSpecies}$. The equation can then be estimated using traditional
Monte-Carlo methods giving
\begin{equation}
\NumInteractions \approx \frac{V}{n} \sum_\NuSpecies \sum_\TargetSpecies \sum_i
\DiffFlux_\NuSpecies(x_i) \, \TargetDensity_\TargetSpecies(x_i) \, \frac{d\sigma_{\NuSpecies\TargetSpecies}}{d\Omega}(x_i) \,,
\end{equation}
where $x_i$ are the selection of the variables of integration for the $i$th point and $V$ is the volume of the space being integrated. The uncertainty on the
integral estimate is thus given by the traditional calculation of the standard deviation. The variance
of the integral estimate can be improved through the use of importance sampling, such as VEGAS~\cite{Lepage:1977sw,Lepage:1980dq}.
The integral over the neutrino fluxes and over the density of the nuclear targets can be obtained in
a similar method, or by using other numerical integration techniques like quadrature. The flux-averaged
cross section can thus be calculated with Monte-Carlo techniques using Eq.~\ref{eq:AvXSec}.

The results of the above Monte-Carlo calculation would produce a set of weighted events that can be used as is, or can be unweighted
through the following procedure. First, the maximum values for each neutrino species and nucleus can be estimated using Monte-Carlo methods ($w^{\rm max}_{\NuSpecies\TargetSpecies}$).
Second, a neutrino species and nucleus is selected according to the probability
\begin{equation}
   P_{\NuSpecies\TargetSpecies} = \frac{w^{\rm max}_{\NuSpecies\TargetSpecies}}{w^{\rm max}}\,,
\end{equation}
where $w^{\rm max}$ is given by the sum of the maximum weight over all neutrino types and nuclei.
Once a neutrino type and nucleus is selected, a set of initial and final state momenta is generated and the integrand for that particular neutrino type and nucleus is calculated. The event can then be
unweighted by performing an accept-reject step using the ratio of weight of the event to the maximum weight calculated for the given neutrino type and nucleus. In other words, the weight of an event
would be given by
\begin{equation}
    \tilde{w_i} =  w^{\rm max} \sum_{\NuSpecies\TargetSpecies} \Theta\left(\frac{w^{\rm max}_{\NuSpecies\TargetSpecies}}{w^{\rm max}}-\sum_{\substack{f'<f,\\j'<j}}\frac{w^{\rm max}_{f'j'}}{w^{\rm max}}-R_1\right) \Theta\left(\frac{w_i}{w^{\rm max}_{\NuSpecies\TargetSpecies}}-R_2\right)\,,
\end{equation}
with $R_1, R_2 \in [0, 1]$ a uniformly distributed random number. This would result in a collection of
events with either weights $w^{\rm max}$ or zero, and the average of these events would give an estimate
of the total flux-averaged cross section, and only the events with non-zero weight are required to be
written out as long as the number of attempted events (\textit{i.e.} the total including the zero weight events)
is also tracked. There are technical issues with directly using the true maximum sampled, since the
maximum depends on the number of samples taken. There are many ways to mitigate this numerical issue, one
approach is discussed in Section 4A of Ref.~\cite{Gao:2020zvv}.

\section{Example Event Graphs}

\begin{figure}
    \centering
    \includegraphics[width=\textwidth]{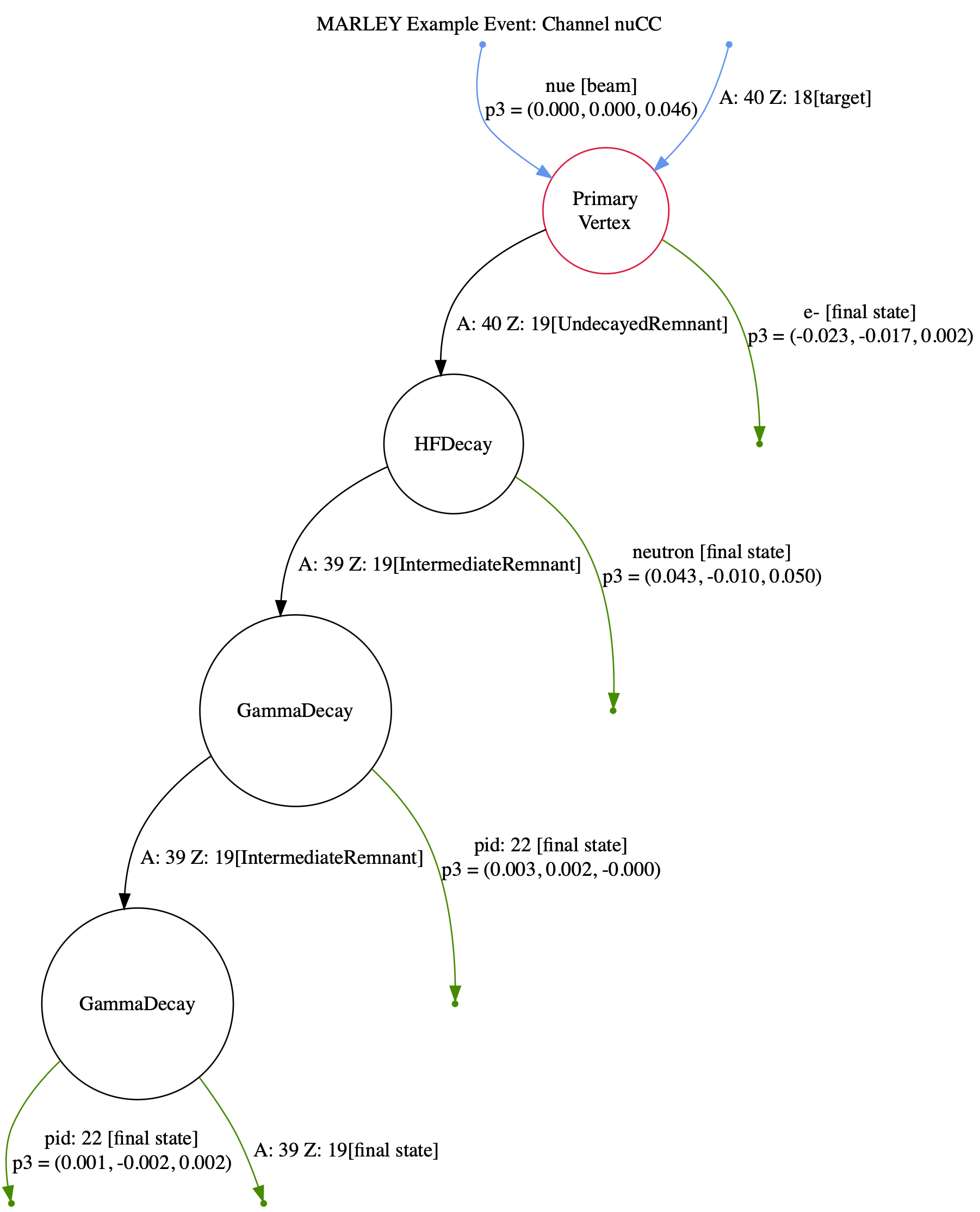}
    \caption{A MARLEY event graph in the NuHepMC format.}
    \label{fig:marleyevg}
\end{figure}

\begin{figure}
    \centering
    \includegraphics[width=0.8\textwidth]{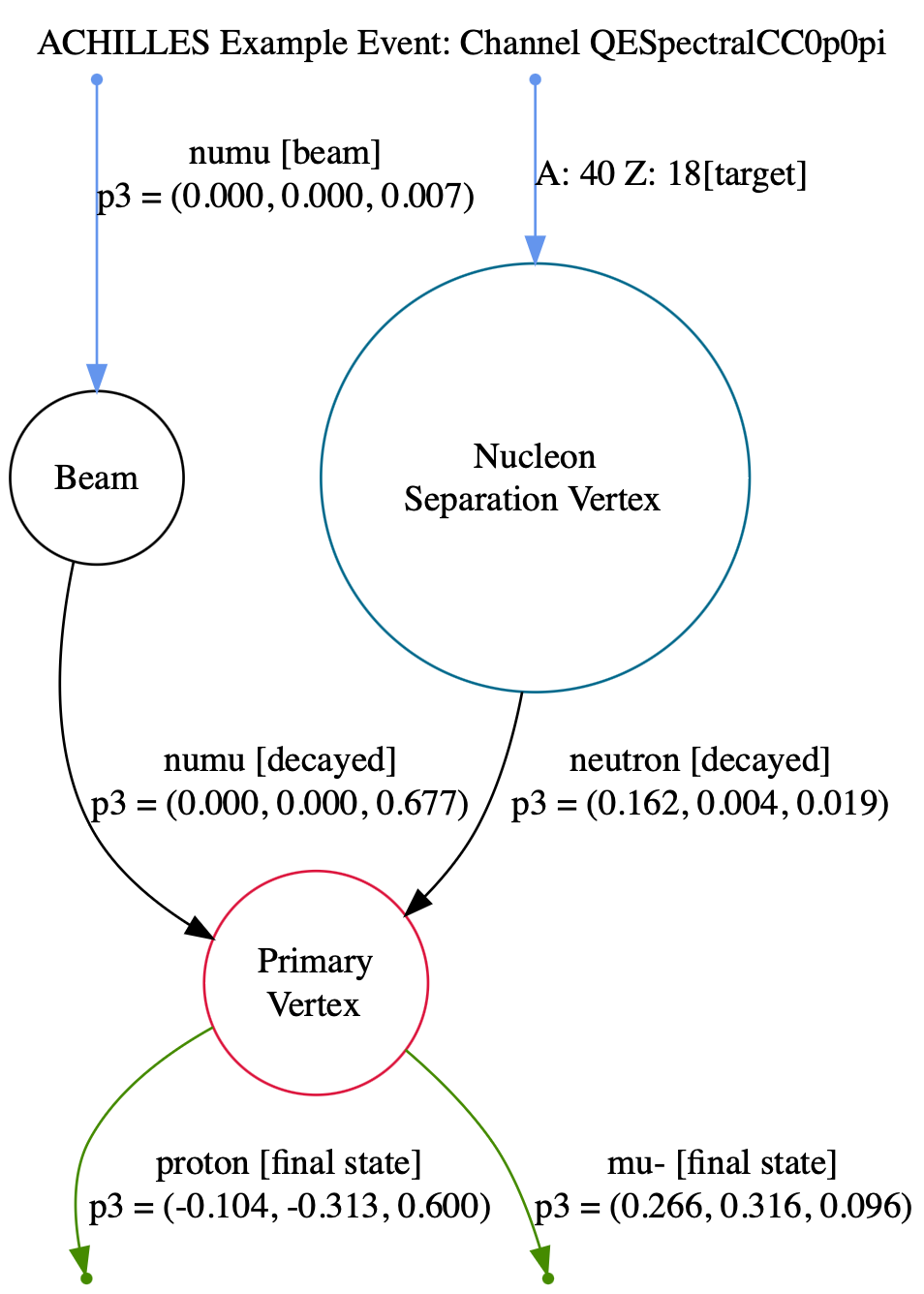}
    \caption{An ACHILLES event graph in the NuHepMC format.}
    \label{fig:achillesevg}
\end{figure}

\begin{figure}
    \centering
    \includegraphics[width=\textwidth]{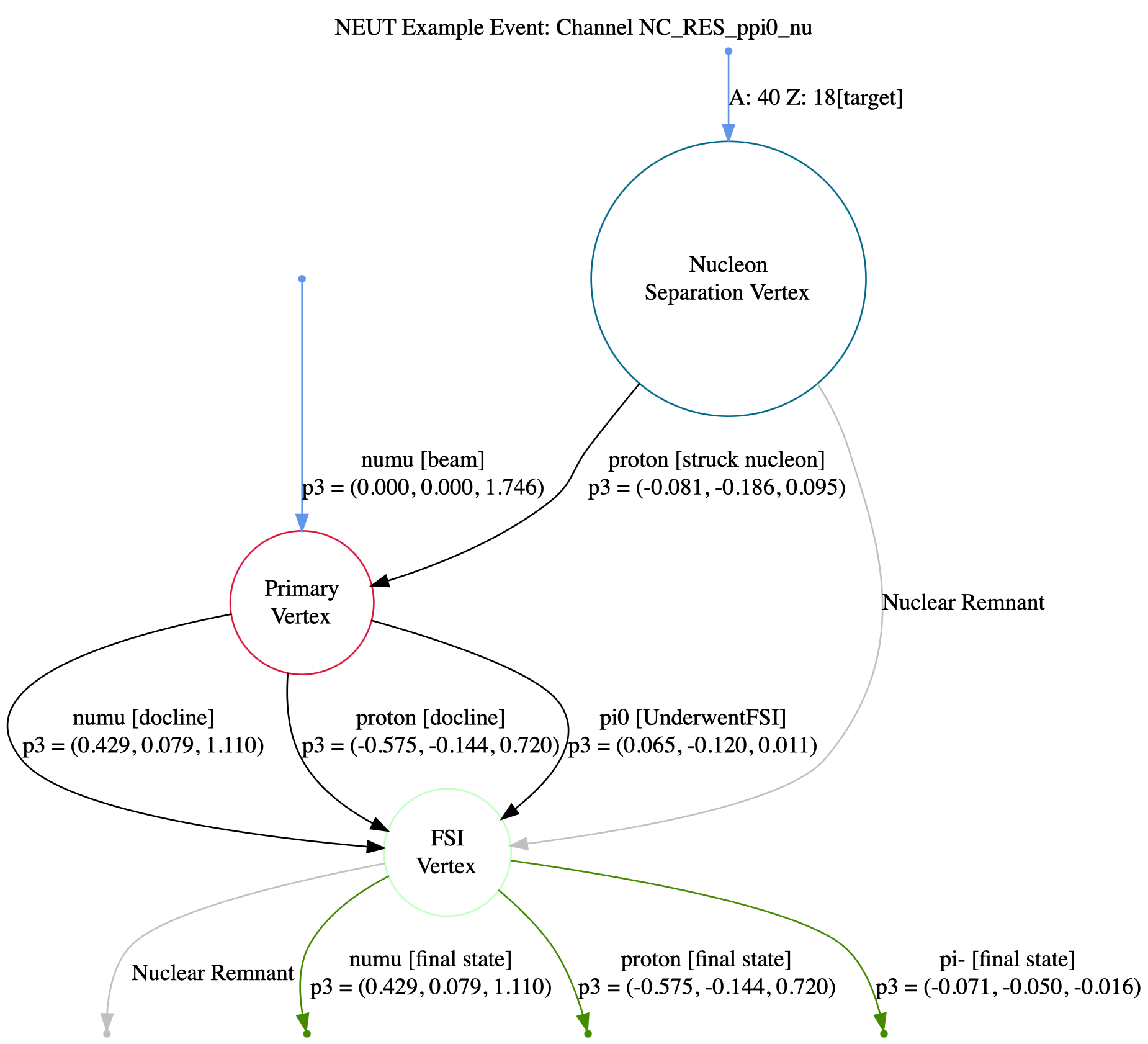}
    \caption{A NEUT event graph in the NuHepMC format.}
    \label{fig:neutevg}
\end{figure}

\begin{figure}
    \centering
    \includegraphics[width=\textwidth]{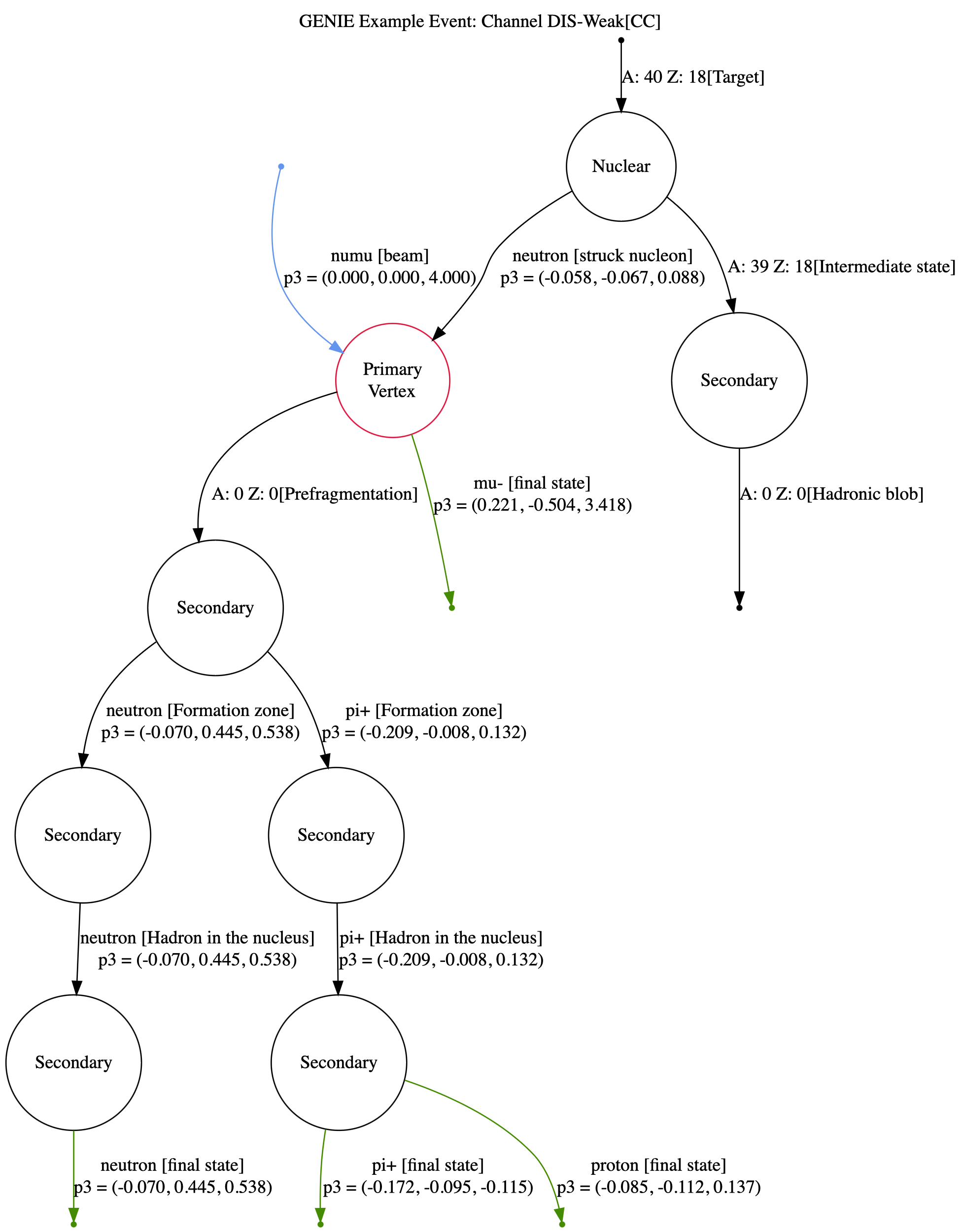}
    \caption{A GENIE event graph in the NuHepMC format.}
    \label{fig:genieevg}
\end{figure}

\begin{figure}
\centering
    \includegraphics[width=\textwidth]{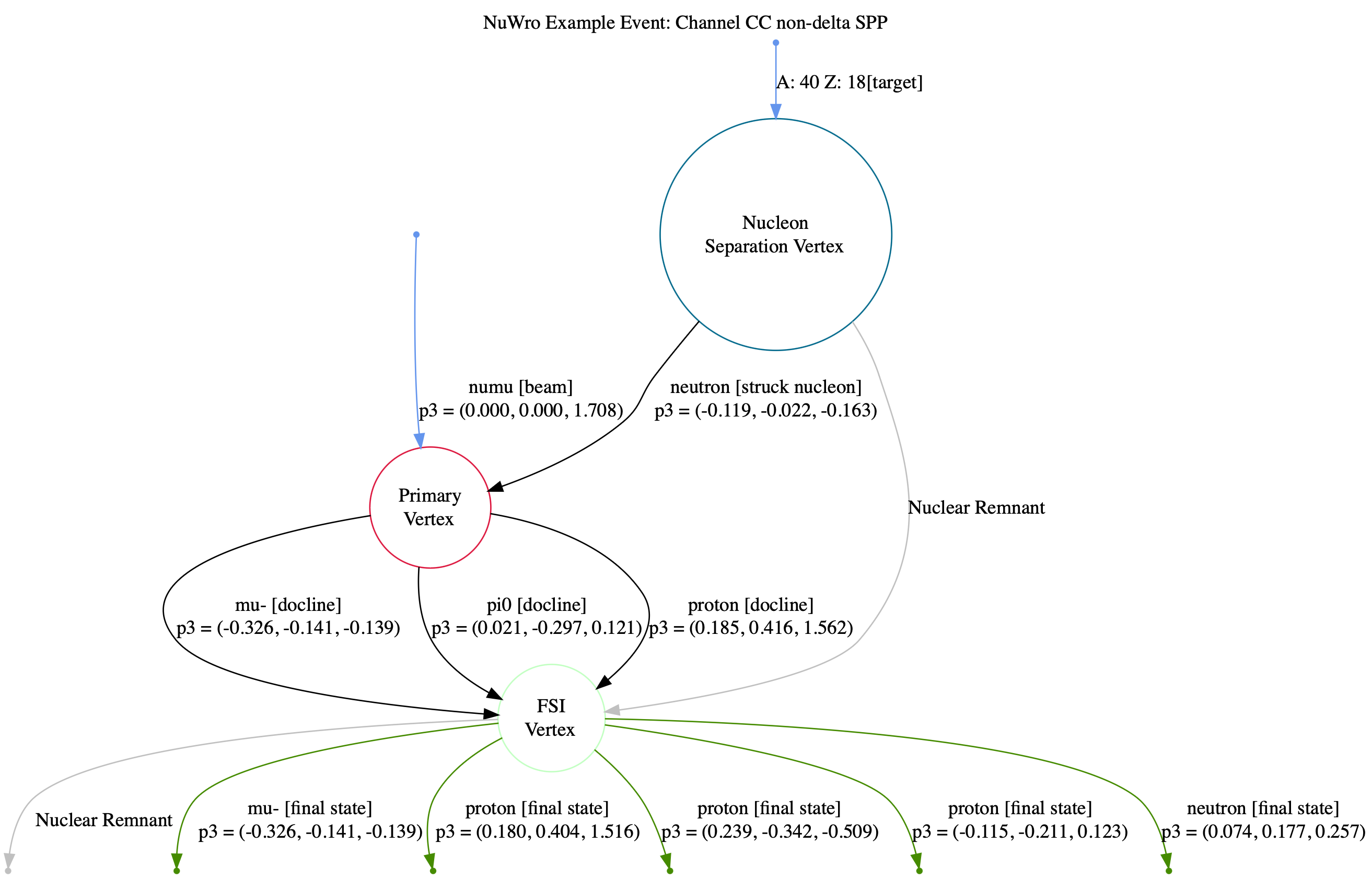}
    \caption{A NuWro event graph in the NuHepMC format. }
    \label{fig:nuwroevg}
\end{figure}

Figs.~\ref{fig:marleyevg},~\ref{fig:achillesevg},~\ref{fig:neutevg},~\ref{fig:genieevg} and~\ref{fig:nuwroevg} show some example event graphs for neutrino event generators that natively implement, or are convertible-to, the NuHepMC format. Although the details of each generator's implementation of this standard differ, the constraints that are imposed enable consistent usage for the most common tasks without any knowledge of each generator's implementation details.


\section{Extracting Bibliography Information}
\label{appendix:bib}
If the NuHepMC file signals that convention \ref{gen:citations} has been used, then it is
possible to run the file through an external tool called \textsc{HEPReference}~\cite{Isaacson_NuDevTools_HEPReference_v0_1_2024}\footnote{The code can be found at: \url{https://github.com/NuDevTools/HEPReference}} to
produce a BibTeX file and a short text blurb showing how to cite all the papers
used to produce the events. An example use case can be found in the repository.

\end{appendix}

\bibliography{biblio}

\end{document}